\documentclass[twocolumn]{aastex6}
\usepackage{natbib}
\usepackage{graphicx}
\usepackage{epstopdf}
\usepackage{amsmath}

\shorttitle{The Magnetic Field of IRDC G28.23}

\newcommand{\Mphi}{M/$\Phi_B$}
\newcommand{\Mphin}{M/$\Phi_{BN}$}
\newcommand{\Mphip}{M$_{\parallel}$/$\Phi_{\perp}$}

\begin{document}
	
	\title{Tracing the Magnetic Field of IRDC \replaced{G28.23}{G028.23-00.19} Using NIR Polarimetry}
	
	\author{Sadia Hoq\altaffilmark{1}, D. P. Clemens\altaffilmark{1}, Andr\'{e}s E. Guzm\'{a}n\altaffilmark{2}, and Lauren R. Cashman\altaffilmark{1}}
	\altaffiltext{1}{Institute for Astrophysical Research, 725 Commonwealth Ave, Boston University, Boston, MA 02215; shoq@bu.edu, clemens@bu.edu, lcashman@bu.edu}
	\altaffiltext{2}{Departamento de Astronom\'{i}a, Universidad de Chile, Camino el Observatorio 1515, Las Condes, Santiago, Chile, aguzman@das.uchile.cl}
	
	\begin{abstract}
		The importance of the magnetic (B) field in the formation of Infrared Dark Clouds (IRDCs) and massive stars is an ongoing topic of investigation.  We studied the plane-of-sky \replaced{magnetic }{B-}field for one IRDC, G028.23-00.19, to understand the interaction between the field and the cloud.  We used near-IR background starlight polarimetry to probe the \replaced{magnetic }{B-}field, and performed several observational tests to assess the \deleted{magnetic} field importance.  The polarimetric data, taken with the Mimir instrument, consisted of \deleted{archival} $H$-band \replaced{polarizations from the Galactic Plane Infrared Polarization Survey (GPIPS) and targeted deep}{and} $K$-band observations, totaling 17,160 stellar measurements.  We traced the plane-of-sky B-field morphology with respect to the sky-projected cloud \replaced{major axis}{elongation}.  We also found the relationship between the estimated B-field strength and gas volume density, and computed estimates of the normalized Mass-to-Magnetic Flux ratio.  \deleted{The cloud volume density was found by fitting a Plummer-like profile to the column density map.}  The B-field orientation with respect to the cloud did not show a preferred alignment, but did exhibit a large-scale pattern.  The plane-of-sky B-field strengths ranged from 10--165 $\mu$G, and the B-field strength dependence on density followed a power law with an index consistent with 2/3.  \added{The Mass-to-Magnetic Flux ratio also increased as a function of density.  The relative orientations and relationship between B-field and density} \deleted{results} imply that the B-field was not dynamically important in the formation of the IRDC.  \added{The increase in Mass-to-Flux ratio as a function of density, though, indicates a dynamically important B-field.  Therefore, it is unclear whether the B-field influenced the formation of G28.23.}  However, it is likely that the presence of the IRDC changed the \added{local} B-field morphology\deleted{ in the region}.
	\end{abstract}
	
	\section{Introduction}\label{introduction}
		Infrared Dark Clouds (IRDCs) are dense (H$_2$ column densities $\sim$10$^{22}$--10$^{23}$ cm$^{-2}$) filamentary structures that host high-mass star formation \citep{Rathborne2006}.  As such, these clouds play important roles in the evolution of the Galaxy.  However, much is still not known about IRDC formation and evolution.  One open problem is the unknown role of magnetic (B) fields in the formation of IRDCs and any potential star formation they host.
		
		While many studies have probed the B-fields of molecular clouds using techniques such as the Zeeman effect, background starlight polarimetry, and dust emission polarimetry \citep[e.g.,][]{Troland2008, Chapman2011, Dotson2010}, studies of the B-fields of IRDCs have mainly focused on the small size scales of cores and clumps \citep[e.g.,][]{Cortes2008, Tang2009, Girart2009, Sridharan2014, Zhang2014}.  Several studies have probed the interstellar medium (ISM) between clouds \citep{Heiles2000, Li2009}, but few have probed the cloud scales (10s of pc) to study the interaction between the B-field and an IRDC as a whole \citep[e.g.,][]{Sugitani2011, Pillai2015}.  
		
		To test whether B-fields play a dominant role in IRDC and massive star formation \citep[e.g.,][]{Nakamura2008}, or are themselves influenced by more dominant forces \citep[e.g.,][]{Padoan1999}, such as turbulence or gravity, B-field properties must be observationally related to other physical cloud properties.  
		
		\replaced{Several observational tests have been used to determine the role of B-fields in molecular cloud and star formation.}{Many models and simulations have studied the formation of filamentary molecular clouds and the forces that control their formation \citep[e.g., ][]{Nakajima1996, Ostriker2001, Padoan2002, Hennebelle2013, Van2014, Li2015}.  Several of these studies \citep[e.g., ][]{Ostriker2001, Van2014, Li2015} considered the role of B-fields in cloud formation to predict the observational signatures of weak and strong B-fields.  One of the most straightforward signatures is the relative orientation of the B-field with respect to the filament orientations.}  For example, the preferential direction of B-fields relative to \deleted{the} cloud orientation can reveal whether material flowed along field lines to create clouds.  If the field is more likely found to be either perpendicular or parallel to the cloud orientation, then the B-field very likely played a role in the cloud's formation \citep{Li2009}.

		Other \added{observational} tests rely on the relative strength of the B-field.  The relative strength of the gravitational potential compared to the local B-field \added{flux} \citep[`Mass-to-Flux' ratio, \Mphi, \added{see review by }][]{Crutcher2012} of a clump reveals the importance of the B-field compared to gravity.  Additionally, if the power-law dependence of the B-field strength on cloud gas volume density is shallower than 2/3, the field likely influenced the flow of material \replaced{in}{during} cloud formation \citep{Li2015}.
		
		Near-Infrared (NIR) background starlight polarimetry provides a way to probe the B-field on scales of $\sim$1 to tens of pc at the distances to most IRDCs.  The polarization signal is caused by aspherical dust grains spinning with their long axes aligned mostly perpendicular to the intervening B-field \citep{Lazarian2007}.  The linear polarization signal imparted on background starlight by the asymmetric dichroic extinction of the dust grains follows the orientation of the B-field in the plane of the sky.  Therefore, the orientations of the NIR polarizations trace the plane-of-sky B-field.  NIR polarimetry can reveal the plane-of-sky B-field morphology over large fields of view and wide ranges of column and volume densities \citep[e.g.,][]{Clemens2012c}.  In addition, the plane-of-sky B-field strength can often be inferred by using the Chandrasekhar-Fermi Method \citep[hereafter CF Method]{CF1953}, which combines the polarization measurements with complementary cloud density and gas velocity information.
		
		\subsection{IRDC G28.23}
\begin{figure*}
				\includegraphics[width = 6in]{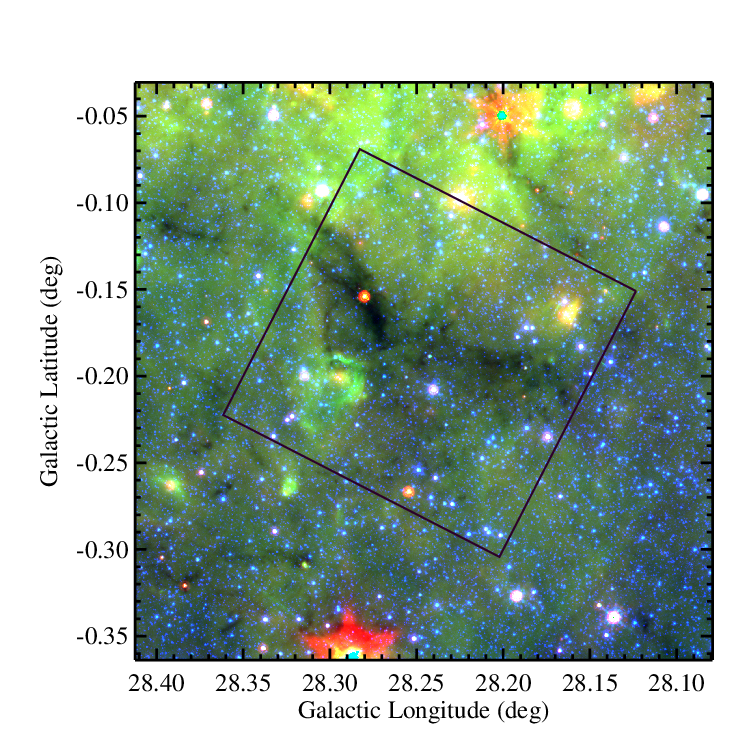}
				\caption{Three-color \added{{\it Spitzer}} GLIMPSE \citep{Benjamin2003} \replaced{+}{and} MIPSGAL \citep{Carey2009} (blue: 3.6, green: 8, red: 24 $\mu$m) image of G28.23.  A black box representing the field of view of the $K$-band polarimetric \replaced{data}{observations} (approximately 10$\times$10 armin) is overlaid.  The bright source at $l$ = 28$\fdg$275, $b$ = $-$0$\fdg$15 is an unrelated foreground star.}
				\label{figglimpsefov}
		\end{figure*}

		In this study, we \replaced{evaluate}{evaluated} the plane-of-sky B-field toward IRDC \replaced{G28.23}{G028.23-00.19 (hereafter referred to as G28.23)} \citep[$l$ = 28$\fdg$23, $b$ = $-$0$\fdg$19;][]{Rathborne2006}, using NIR polarimetric observations, to determine \replaced{its}{the} role \added{of the B-field} in the formation of the IRDC.  Figure \ref{figglimpsefov} shows \replaced{the}{a} 3-color GLIMPSE\replaced{+}{ and} MIPSGAL \citep{Benjamin2003, Carey2009} image of G28.23.  \added{Both GLIMPSE (Galactic Legacy Infrared Mid-Plane Survey Extraordinaire) and MIPSGAL (MIPS Galactic Plane Survey) are infrared surveys of the Galaxy conducted with the {\it Spitzer} Space Telescope.  GLIMPSE observations were taken in four bands: 3.6, 4.5, 5.8, and 8 $\mu$m, and MIPSGAL observations were taken in 24 and 70 $\mu$m.}  The IRDC can be seen as the dark extinction feature against the bright background emission\footnote{The bright IR point source at the top left of the cloud is an unassociated OH/IR foreground star, at a LSR radial velocity of $\sim$52 km~s$^{-1}$ \citep{Bowers1989}, which is different from the 80 km~s$^{-1}$ velocity of G28.23 \citep{Sanhueza2013}.}.  \deleted{Located at a distance of 5.1 kpc, G28.23 is a dense, quiescent IRDC that hosts a massive starless clump (Sanhueza et al. 2013).}    \replaced{G28.23}{Located at a distance of 5.1 kpc, G28.23 is a dense, quiescent IRDC that} hosts one of the most massive quiescent cores found in an IRDC, and is a likely candidate to host future high-mass star formation.  Because of its quiescent nature, G28.23 is an ideal laboratory in which to study the interaction between \replaced{a massive, quiescent IRDC and the surrounding B-field.}{an IRDC and the surrounding B-field.  With no signs of active star formation, G28.23 provides an opportunity to study the interaction between an IRDC and the surrounding B-field before any potential disruption by active star formation.}
		
		\subsection{Methodology}\label{methodology}
		We pursued answers to four questions regarding the plane-of-sky B-field, as revealed by deep NIR observations, to ascertain the field's importance in the formation of G28.23.  
		
		1. Do polarization percentages, as probed by NIR background starlight polarimetry, increase as a function of extinction?  If the polarization percentages do not increase with extinction, \replaced{then the NIR polarization measurements only probe the skin of the cloud (Arce et al. 1998)}{then it is possible that the NIR polarization measurements only probe the skin of the cloud \citep{Arce1998} and do not reveal the B-field properties deeper into the cloud.  Depolarization along the line of sight, where different layers of material exhibit different polarization orientations that cancel when summed, can also cause this effect.  If the polarization increases with extinction, then this is evidence that the material in G28.23 positively increases the measured polarization percentage.  Therefore, the B-field can be probed by NIR polarization of background stars.}
		
		2. Is the plane-of-sky B-field preferentially aligned with the \added{major axis} orientation of G28.23?  Because the angle of the cloud major axis orientation with respect to plane of the sky is not known, we will use the cloud elongation as a proxy for the cloud's major axis.  If the B-field is perpendicular or parallel to the cloud \added{elongation}, then \replaced{it}{the field} likely played a role in the cloud formation.  Otherwise, if the field is randomly oriented with respect to the cloud, the field likely did not strongly influence cloud formation.
		
		3. What is the power-law dependence between B-field strength and cloud volume density?  \deleted{If the power law index is shallower than 2/3, the field may have played a role in collecting material.}  A weak B-field frozen into isotropically collapsing material would follow a power law of 2/3 \citep{Crutcher2010}, \added{whereas a strong field would follow a power law shallower than 2/3}.  
		
		4. How does \Mphi~vary across the cloud?  While NIR polarimetry will not probe the densest cloud interiors, it will probe the B-field in the outer \added{to middle} cloud \replaced{layers}{regions}. \replaced{These observations}{The \Mphi~estimates} will reveal the relative changes of the B-field strength with respect to gravity in the outer layers of G28.23 \added{, especially in the density ranges where ambipolar diffusion is predicted to operate \citep[e.g., ][]{Mouschovias1979}}.

		Using a combination of new and archival NIR polarimetry, archival submm and far-IR (FIR) dust emission, and published molecular line data, we studied the magnetic and physical properties of G28.23.  The paper is organized as follows.  Section \ref{observations} describes the observations and archival data products used.  Section \ref{analysis and results} outlines the data analysis and results.  Section \ref{Discussion} discusses the implications of the results on the importance of the B-field, and Section \ref{summary} summarizes the study.

	\section{Observations}\label{observations}
		\subsection{NIR Polarimetry}
			NIR polarimetric observations in both $H$-band (1.6 $\mu$m) and $K$-band (2.2 $\mu$m) of G28.23-00.19 were obtained using the Mimir instrument \citep{Clemens2007} on the 1.8m Perkins Telescope in Flagstaff, AZ.  The instrument field of view (FOV) was 10x10 arcmin with a plate scale of 0.58 arcsec per pixel.  Mimir used a compound half-wave plate (HWP) in conjunction with a fixed, cold wire-grid.  The $H$-band data were taken from Data Release~2 of the Galactic Plane Infrared Polarization Survey \citep[GPIPS,][]{Clemens2012c}\footnote{http://gpips0.bu.edu/Data\_Release/}.  GPIPS spanned 76 sq. degrees of the inner Galaxy, from 18--56$^{\circ}$ in Galactic longitude and $\pm$1$^{\circ}$ in latitude, which fully covered the location and extent of G28.23.  The survey region consists of 3,237 individual pointings, each covering a 10x10 arcmin area.  Each GPIPS observation consisted of 96 images (one each at 16 unique HWP positions at six dither positions on the sky) with an exposure time of 2.5 seconds.  The total integration time of each observation was $\sim$4 minutes.  Sky conditions were clear, and each GPIPS field was required to meet a 2 arcsec seeing criterion.
			
			The $H$-band data for a 20$\times$20 arcmin region (the region shown in Figure \ref{figglimpsefov}) covering the IRDC, as well as its environment, were extracted from the GPIPS database.  This region is aligned in Galactic coordinates, with center ($l$, $b$) = (28.244, $-$0.200).
			
			Targeted $K$-band observations, covering about 10.4$\times$10.8 arcmin, centered on coordinates $l$ = 28$\fdg$247 and $b$ = $-$0$\fdg$19, and aligned in R.A. and decl., were obtained over five nights in 2013 September and 2014 June.  These consisted of 14 separate observations, each with 96 individual 15 second exposures.  The total integration time was 5.6 hours.  These deep $K$-band observations probed the more extincted regions, while the $H$-band data covered a larger FOV.  The $K$-band sky coverage is outlined as the black box in Figure~\ref{figglimpsefov}.  Hereafter, references to the 10$\times$10 arcmin FOV centered on the cloud refer to this 10.4$\times$10.8 arcmin region.
			
			The calibration of the NIR polarimetric data is described in \citet{Clemens2012b}.  The data were reduced using the custom IDL packages Mimir Software Package Basic Data Processing (MSP-BDP) and Photo POLarimetry (MSP-PPOL) \citep{Clemens2012b}.  The main reduction and processing steps included taking dome flat fields at each HWP position to correct for variation across the Mimir FOV, accounting for instrumental polarization by observing globular cluster stars, and converting the instrumental polarization position angles to Equatorial coordinates via observations of polarimetric standard stars from \citet{Whittet1992}.  The resulting combined polarimetric catalog contains the properties of individual stars down to magnitudes of $\sim$13 for $H$-band and $\sim$14-14.5 for $K$-band, as measured by Mimir.  The polarization position angles, PAs, are measured as the angle East of North in Equatorial coordinates, and can be transformed to Galactic coordinates (Galactic PAs or GPAs) by adding 62.8$^{\circ}$.
		
		\subsection{Additional Datasets}
			In addition to NIR polarimetry, we used NIR and mid-IR (MIR) photometry from the 2MASS \citep{Skrutskie2006}, UKIDSS \replaced{(Lawrence et al. 2007)}{\citep[UKIRT Infrared Deep Sky Survey; ][]{Lawrence2007}}, and GLIMPSE \citep{Benjamin2003} catalogs, dust continuum data from the {\it Herschel} infrared Galactic Plane Survey (Hi-GAL) \citep{Molinari2010} and the APEX Telescope Large Area Survey of the Galaxy (ATLASGAL) \citep{Schuller2009}, and molecular line data from the $^{13}$CO Galactic Ring Survey \citep[GRS; ][]{Jackson2006}.  The NIR and MIR photometry were used to estimate the extinctions to the polarization stars.  The dust continuum data were used to create an H$_2$ column density map of the cloud.  The $^{13}$CO data, which \replaced{has}{have} an angular resolution of 46 arcsec, were used to find the gas velocities and line widths in the less dense regions of the cloud.
			
	\section{Analysis and Results}\label{analysis and results}
		Because the B-field of \replaced{the cloud}{G28.23} can only be probed by stars lying beyond it, we first found the polarimetric stars that were background to the cloud.  The relative locations of the stars with respect to the cloud (background, foreground), were determined by comparing the stellar extinctions to the dust emission-traced cloud extinction.  We also accounted for foreground extinction and polarizing layers.  The column density map of the cloud was found by using the {\it Herschel} Hi-GAL and ATLASGAL dust emission data.  The \added{background stellar} polarization PA orientations were found and compared to the cloud orientation.  The cloud volume density was derived from the column density, and used in estimating the plane-of-sky B-field strength.  These steps are described in more detail below.
	
		\subsection{Catalog of NIR Polarimetric and Photometric Stars}\label{catalog of nir stars}
		  A list of NIR stars was created using the 2MASS and UKIDSS photometric data within the $20\times20$~arcmin region of Figure \ref{figglimpsefov}.  Stars with $H$-band magnitudes brighter than 13$^{th}$ mag were selected from 2MASS, and stars fainter than 13$^{th}$ mag were selected from UKIDSS.  Only stars with $H$-band uncertainties less than 0.3 mag were retained.  These stars were then matched to the \added{Mimir} $H$ and $K$-band polarization catalogs.  The number of 2MASS+UKIDSS stars within the 20$\times$20 arcmin FOV with polarizations in either $H$, $K$, or both was 17,160.  Of these, 3,280 stars had both $H$ and $K$-band polarization entries ($H$-pol, $K$-pol), while 12,554 stars were found only in $H$-pol and 1,326 were found only in $K$-pol.  The number of $H$-pol entries is much higher because \replaced{it spans}{the $H$-pol data span} a larger area \added{and are located throughout less extincted regions in the FOV than the $K$-band targeted observations}.  This \added{NIR} catalog of stars was then matched to the \added{MIR} GLIMPSE catalog.  Of the 17,160 \added{NIR} stars, 12,490 were positionally-matched to GLIMPSE 4.5~$\mu$m point sources.
		
		Table \ref{table1} lists the polarimetric properties of the stars - the polarimetric bands in which \replaced{they have entries}{the stars were observed}, polarization percent, Galactic polarization PAs (GPAs), and Equatorial Stokes Q and U parameters (Q$_E$, U$_E$), along with uncertainties.  The reported GPAs are measured from the North Galactic Pole with GPAs increasing along the East of North direction, where 90$\degr$ is parallel to the Galactic plane and 0 and 180$\degr$ are perpendicular to the plane.  These GPAs were derived by rotating the Equatorial PAs computed from Stokes U$_E$ and Q$_E$ parameters.  Table \ref{table2} lists the photometric properties of these stars in the same order, including the 2MASS or UKIDSS designation, stellar colors and magnitudes, and the relative extinctions and distances (discussed below).  
		
		The polarization percentages reported and used in the analysis have been Ricean corrected \citep{Wardle1974} to account for positive bias\added{, where $\sigma_P$ is equal to the uncertainty in the polarization percentage}: 
		
		\begin{align}
		P_{uncorrected} = \sqrt{U_E^2 + Q_E^2},\\
		P_{corrected} = \sqrt{P_{uncorrected}^2~-~\sigma_{P}^{2}}.
		\end{align}
		All polarization measurements reported in Table \ref{table1} and in the results are P$_{corrected}$.  Stars with \replaced{polarization}{P$_{corrected}$} signal-to-noise ratios (SNR) greater than 2.5 and \deleted{polarization} uncertainties less than 5\% were classified as ``high SNR'' stars.  \added{In the majority of the following analysis, only high SNR stars are used.  This selection criteria eliminates faint stars and bright stars with little to no polarization percentages and low uncertainties.}

		\floattable
		\begin{deluxetable}{cccccccccccc}
			\tabletypesize{\tiny}
			\tablecolumns{12}
			\tablecaption{Stellar Polarimetric Properties of G28.23 Field of View}
			\tablewidth{0pt}
			\colnumbers
			\rotate
			\tablehead{
				\multicolumn{8}{c}{~~~~~~~~~~~~~~~~~~~~~~~~~~~~~~~~~~~~~~~~~~~~~~~~~~~\underline{~~~~~~~~~~~~~~~~$K$-Band Polarization~~~~~~~~~~~~~~~~~~~~~~~~~~~}} & 
				\multicolumn{4}{c}{\underline{~~~~~~~~~~~~~~~~~~$H$-Band Polarization~~~~~~~~~~~~~~~~~}} \\
				\colhead{Star} &
				\colhead{$l$}  &
				\colhead{$b$}  &
				\colhead{Band\tablenotemark{a}} &
				\colhead{P\tablenotemark{b}\tablenotemark{c}} &
				\colhead{GPA} & 
				\colhead{Q$_E$} &
				\colhead{U$_E$} &
				\colhead{P} &
				\colhead{GPA} &
				\colhead{Q$_E$} &
				\colhead{U$_E$} \\
				\colhead{Number} &
				\colhead{($^{\circ}$)} &
				\colhead{($^{\circ}$)} &
				\colhead{Match}  &
				\colhead{(\%)}   &
				\colhead{(deg)}  &
				\colhead{(\%)}     &
				\colhead{(\%)}     &
				\colhead{(\%)}   &
				\colhead{(deg)}  &
				\colhead{(\%)}     &
				\colhead{(\%)}     
			}
			\startdata
			4216 &    28.1559  &     -0.1609 & HK & 1.6 (0.9) & 142 (15) & $-$1.7 (0.9) & 0.7 (0.7) & 2.3 (3.6) & 139 (44) & $-$3.8 (3.5) & 2.0 (3.9) \\
			4217 &    28.1559  &     -0.0923 & H  & $\cdots$    & $\cdots$ & $\cdots$ & $\cdots$ & 7.9 (8.6) & 117 (31) & $-$3.7 (6.3) & 11.0 (8.8) \\
			4218 &    28.1560  &     -0.1861 & K  & 57 (93) & 122 (47) & $-$51 (99.9) & 96 (63) & $\cdots$ & $\cdots$ & $\cdots$ & $\cdots$ \\
			4221 &    28.1560  &     -0.2120 & HK & 10 (15) & 100 (42) & 5 (22) & 18 (15) & 6.7 (4.0) & 100 (17) & 2.0 (4.3) & 7.6 (4.0) \\
			4222 &    28.1560  &     -0.3196 & H  & $\cdots$ & $\cdots$ & $\cdots$ & $\cdots$ & 1.4 (0.7) & 88 (14) & 1.0 (0.7) & 1.2 (0.7) \\
			4223 &    28.1561  &     -0.0741 & H  & $\cdots$ & $\cdots$ & $\cdots$ & $\cdots$ & 8.6 (3.4) & 94 (11) & 4.3 (3.2) & 8.2 (3.4) \\
			\enddata
			\label{table1}
			\tablenotetext{a}{H: Entry only in $H$-pol, K: Entry only in $K$-pol, HK: Entries in both $H$ and $K$-pol.}
			\tablenotetext{b}{Uncertainties of quantities are listed in parentheses.}
			\tablenotetext{c}{The polarization measurements listed are have been corrected for bias, as described in the text.}
			\tablecomments{This table is available in its entirety in a machine-readable form in the online journal.  A portion is shown here for guidance regarding its form and content.}
			\tablecomments{Missing data are shown here as ellipses, but are represented by the number 9999.99 in the online table.}
		\end{deluxetable}
		
		
		\floattable
		\begin{deluxetable}{ccccccccccc}
			\tabletypesize{\tiny}
			\tablecolumns{11}
			\tablecaption{Stellar Photometric Properties of G28.23 Field of View}
			\tablewidth{0pt}
			\colnumbers
			\rotate
			\tablehead{
				\colhead{Star} &
				\colhead{NIR\tablenotemark{a}}  &
				\colhead{J}    &
				\colhead{H}    &
				\colhead{K}    &
				\colhead{4.5 $\mu$m\tablenotemark{b}} &
				\colhead{H-K}  &
				\colhead{H-4.5 $\mu$m} &
				\colhead{A$_V$} &
				\colhead{A$_V$} &
				\colhead{Relative\tablenotemark{c}} \\
				\colhead{Number} &
				\colhead{Desig.} &
				\colhead{(mag)}  &
				\colhead{(mag)}  &
				\colhead{(mag)}  &
				\colhead{(mag)}  &
				\colhead{(mag)}  &
				\colhead{(mag)}  &
				\colhead{(mag)}  &
				\colhead{Method} &
				\colhead{Distance}
			}
			\startdata
			4216 & 438563875392 & 16.150 (0.008) & 13.390 (0.002) & 11.996 (0.001) & 11.00 (0.06) & 1.394 (0.002) & 2.39 (0.06) & 18.7 (0.5) & H-4.5 & B \\
			4217 & 438563881028 & 15.411 (0.005) & 14.777 (0.005) & 14.424 (0.010) & $\cdots$     & 0.353 (0.011) & $\cdots$    & 3.6 (0.2) & H-K & F \\
			4218 & 438563873228 & 16.875 (0.015) & 14.794 (0.005) & 13.537 (0.005) & 12.53 (0.26) & 1.256 (0.007) & 2.26 (0.26) & 17.6 (2.1) & H-4.5 & B \\
			4221 & 438563870876 & 15.200 (0.004) & 13.656 (0.002) & 12.931 (0.003) & 12.23 (0.16) & 0.725 (0.003) & 1.42 (0.16) & 10.9 (1.3) & H-4.5 & B \\
			4222 & 18435111-0423439 & 13.20 (0.04) & 11.58 (0.04) & 10.99 (0.03) & 10.45 (0.06) & 0.60 (0.05) & 1.13 (0.07) & 8.5 (0.6) & H-4.5 & B \\
			4223 & 438563882475 & 15.362 (0.004) & 13.748 (0.002) & 12.940 (0.003) & 12.46 (0.11) & 0.807 (0.003) & 1.29 (0.11) & 9.8 (0.9) & H-4.5 & U \\
			\enddata
			\label{table2}
			\tablenotetext{a}{Stellar designations are those listed in the 2MASS/UKIDSS catalogs.  Stars brighter than 13$^{th}$ mag in $H$-band were taken from 2MASS, and stars fainter than 13$^{th}$ mag were taken from UKIDSS.}
			\tablenotetext{b}{4.5 $\mu$m photometry was retrieved from the GLIMPSE catalog.}
			\tablenotetext{c}{Relative locations of stars with respect to the cloud along the line of sight.  B: Background Star, F: Foreground Star, U: Unknown}
			\tablecomments{This table is available in its entirety in a machine-readable form in the online journal.  A portion is shown here for guidance regarding its form and content.}
			\tablecomments{Missing data are shown here as ellipses, but are represented by the number 99.99 in the online table.}
		\end{deluxetable}
		
\begin{figure} 
			\begin{center}
				\includegraphics[width = \columnwidth]{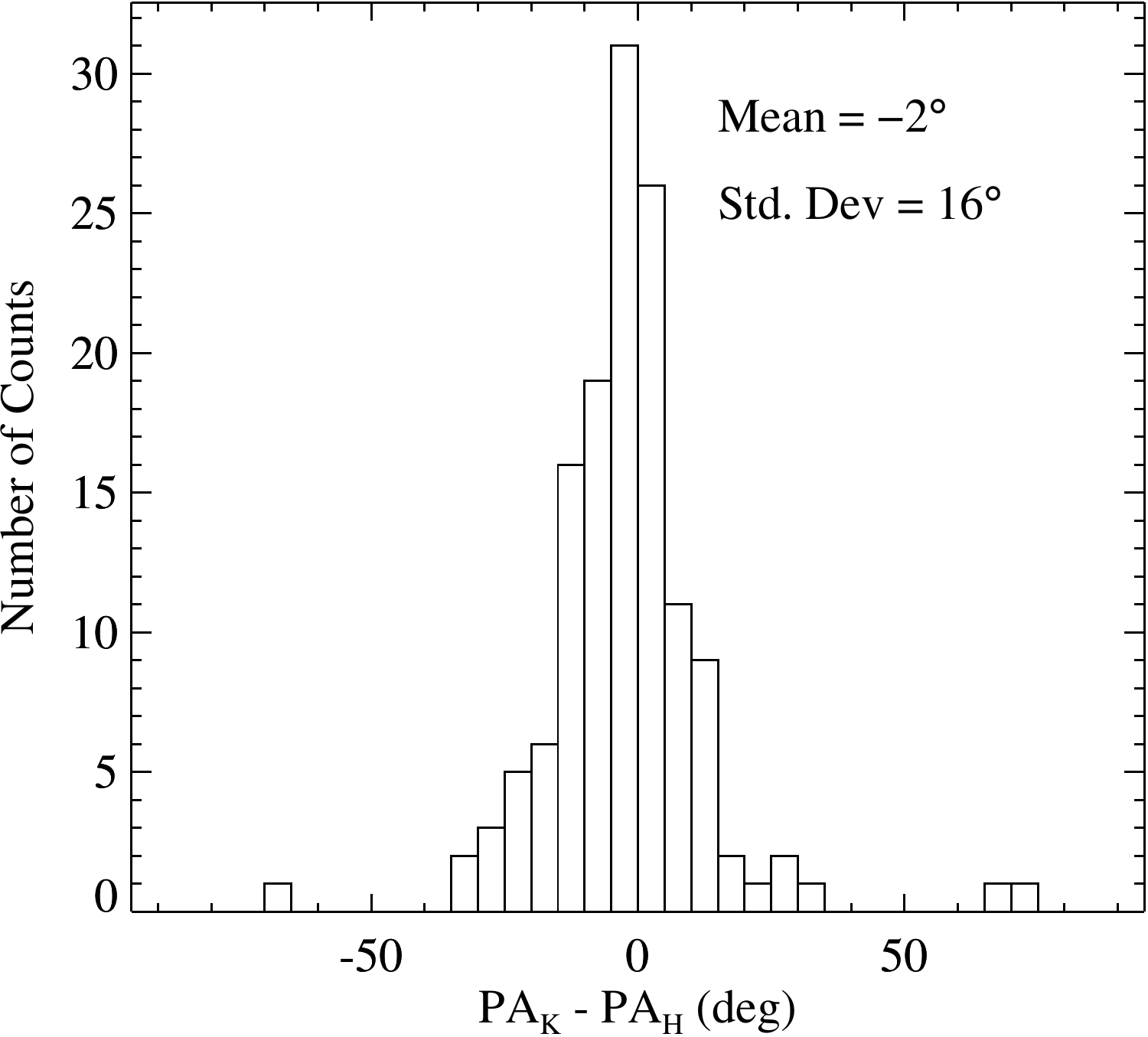}
				\caption{Distribution of the $\delta$PA$_{K-H}$ of high SNR stars.  The variance-weighted mean and \added{weighted} standard deviation of the distribution are $-$2$\degr$ and 16$\degr$, respectively.}
				\label{figdelpahk}
			\end{center}
		\end{figure}

		
		Because many of the stars in the catalog had both $H$ and $K$-pol entries, it was possible to compare the difference in PAs between the two wavelengths ($\delta$PA$_{K-H}$) for each star.  Of the stars with both $H$ and $K$-pol matches, 137 showed high SNR in both bands.  Figure \ref{figdelpahk} shows the distribution of $\delta$PA$_{K-H}$ of these high SNR stars.  The variance-weighted mean $\delta$PA$_{K-H}$ is $-$\replaced{2}{2.1$\pm$0.6}$\degr$ \added{\citep[this uncertainty estimate includes the observational uncertainty of the weighted mean, 0.23$\degr$, added in quadrature to the systematic uncertainty of the Mimir PA measurements of 0.6$\degr$,][]{Clemens2012b}}, with a weighted standard deviation of 16$\degr$.  The standard deviation is close to the uncertainty in PA of a star with polarization signal-to-noise of 2.5 (11.5$\degr$).  Therefore, the measured $H$ and $K$-pol PAs are judged to be identical to within their uncertainties.
		
		\subsection{Selecting Background Stars}
			To probe the B-field of G28.23, it was necessary to select stars that are background to the cloud.  Foreground stars probe material between the cloud and the observer, and therefore, do not probe the cloud B-field.  Finding distances to field stars, however, is difficult using only photometric information.  Therefore, for the purpose of finding stars background to G28.23, we compared the photometrically determined stellar extinctions to the thermal dust emission-derived cloud column density (converted to an extinction map) along the same lines of sight.  Stars exhibiting larger extinctions than the \added{corresponding} cloud extinction were selected as background stars.
			
			\subsubsection{Stellar Extinctions}\label{stellarextinctions}
				\added{The NIR and MIR photometric properties of the polarization stars were used to estimate stellar extinctions.}  Each polarimetric star in Table \ref{table1} contains a photometric entry in Table \ref{table2}.  Not all entries contain a match to the GLIMPSE catalog, as stated in Section \ref{catalog of nir stars}.  \added{The intrinsic range of colors for stars at these wavelengths (on the Rayleigh-Jeans tail) is narrow.  Therefore, colors observed in excess of the intrinsic values of the stars can be reliably attributed to interstellar extinction.}  The extinctions of the stars that were matched to GLIMPSE were estimated via the Rayleigh-Jeans Color Excess \citep[RJCE, ][]{Majewski2011} method if their 4.5~$\mu$m magnitude uncertainties were less than 0.3 mag.  The RJCE method uses the NIR $H$-band and MIR 4.5~$\mu$m magnitudes of stars to determine their color excesses, E($H-4.5$~$\mu$m), following the extinction law of \citet{Indebetouw2005}.  We adopted an intrinsic ($H-4.5$~$\mu$m)$_{\circ}$ equal to 0.08~mag \citep{Majewski2011} for all stars.  For stars that could not be matched to the GLIMPSE catalog, or were matched but their 4.5~$\mu$m magnitude uncertainties were greater than 0.3~mag, we used the Near-Infrared Color Excess \citep[NICE, ][]{Lada1994} method,  which uses the E$(H-K)$ color excesses of stars to estimate their extinctions.  We adopted an intrinsic $(H-K)_{\circ}$ of 0.13~mag for these stars.				  
				
			\subsection{Cloud Extinctions}\label{cloudextinctions}
				We used public {\it Herschel} data from the Hi-Gal project \citep{Molinari2010} and data from ATLASGAL \citep{Schuller2009} to estimate the cloud column density.  {\it Herschel} data for G28.23 are identified with the obs-ID numbers 1342218694/5 and were observed during April 2011 using the \added{SPIRE+PACS} parallel mode, \replaced{observing}{which observed in} five wavebands simultaneously (70, 160, 250, 350, and 500~$\mu$m). The ATLASGAL data provide additional sub-mm measurements at 870~$\mu$m.
				
				The data were reduced and processed following the procedure described in detail in \citet{Guzman2015}.  All maps were convolved to the resolution of the 500~$\mu$m maps ($37$~arcsec) and projected to a common pixel grid.  To study filamentary IRDCs, it is important to subtract the diffuse far-IR emission of the Galactic plane that is not associated with the filament. \added{A background image, constructed for each Hi-GAL field by smoothing the field image, was subtracted from each field.}  The subtraction procedure is described \added{in detail} in \citet{Guzman2015} and is similar to that used in other IRDC studies performed using {\it Herschel} data \citep[e.g.,][]{Battersby2011}.  At 500~$\mu$m, the diffuse component subtracted around G28.23 amounts to 250--300 MJy~sr$^{-1}$, which is comparable to the emission expected from the IRDC itself.
				
				Dust column density and temperature maps were obtained by fitting a single temperature gray-body model to \added{the multiple wavelength intensities of} each pixel.  \replaced{and}{These densities and temperatures} were converted to gas column densities using theoretical dust opacity curves \citep{Ormel2011}. The dust model used contained no ice coatings and had 3$\times$10$^{4}$ years of coagulation for silicate-graphite grains, \added{as might be appropriate for the outer and middle regions of such a dense filament}.  The uncertainties of the dust temperatures and column densities are given in \citet{Guzman2015}, and are on the order of 10\%.  Following \citet{Heiderman2010}, the gas column densities were converted to extinctions via the relation A$_V$~[mag] = N$_{H_2}$~[cm$^2$]/1.37$\times$10$^{21}$~[cm$^2$/mag], where A$_V$ = 1.086~C$_{EXT}$ \citep{Draine2003}, R$_V$ is equal to 5.5, and C$_{EXT}$ at $V$-band is equal to 6.715$\times$10$^{-22}$ cm$^2$~H$^{-1}$ \citep{Weingartner2001}\footnote{http://www.astro.princeton.edu/~draine/dust/dustmix.html.}.  
				
\begin{figure*} 
					\begin{center}
						\includegraphics[width = 6in]{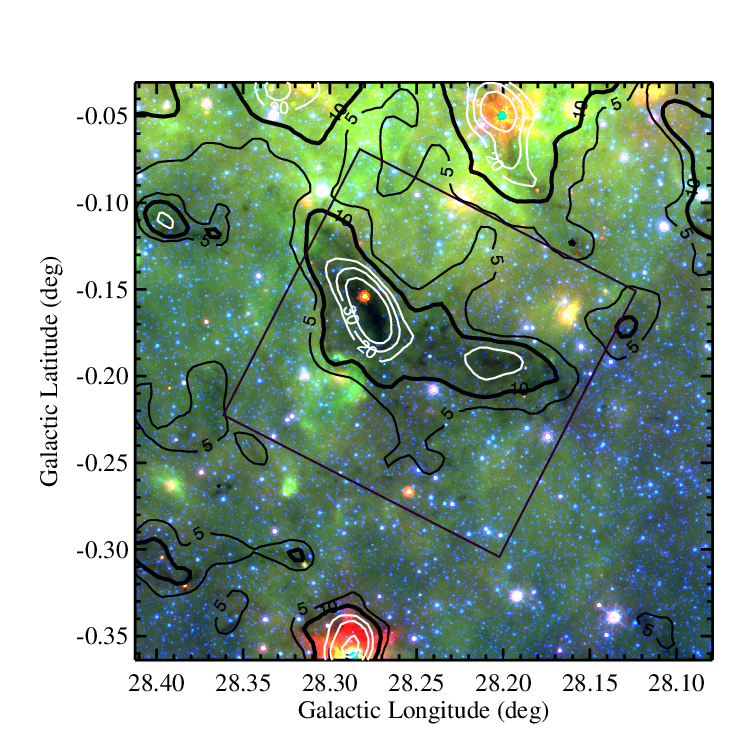}
						\caption{GLIMPSE and MIPSGAL 3-color (blue: 3.6, green: 8, red: 24 $\mu$m) background image of G28.23, with column density contours converted to A$_V$ (levels = [5, 10, 20, 30, 50] mag) overlaid in black and white.  The thick black contour of A$_V$ = 10 mag \deleted{roughly} corresponds to N$_{H_2}$ = \replaced{1}{1.37}$\times$10$^{22}$~cm$^{-2}$.  The IRDC appears as the dark extinction feature in the center of the contours.}
						\label{figglimpsecold}
					\end{center}
				\end{figure*}

				Figure \ref{figglimpsecold} shows the GLIMPSE and MIPSGAL 3-color image of G28.23, with column density-based A$_V$ contours overlaid.  An A$_V$ of 10~mag \deleted{roughly} corresponds to an H$_2$ column density of \replaced{1}{1.37}$\times$10$^{22}$~cm$^{-2}$.  The dark extinction feature within the 50 mag A$_V$ contour in Figure \ref{figglimpsecold} corresponds to the densest region of the IRDC.  The long axis of the cloud, including both the densest regions of the cloud between longitudes 28$\fdg$25 and 28$\fdg$32 and the less dense filament at longitudes less than 28$\fdg$25, extends approximately 12 arcmin.  Due to the proximity of G28.23 to the Galactic mid-plane, it is located in a region of non-negligible, and variable, extinction.  At latitudes below that of the cloud, farther from the mid-plane, the extinction decreases, whereas at latitudes closer to the midplane, the extinction increases.
				
			\subsubsection{Accounting for Foreground Extinction and Polarization}
				Because the dust emission-based cloud extinction estimates accounted for foreground and background extinction \added{unrelated to the IRDC} through removal of \added{surrounding} diffuse emission, a similar type of correction needed to be done for the stellar extinctions prior to the assignment of stellar locations.  To estimate the foreground \added{stellar} extinction, we examined the extinctions of the polarization catalog stars that were spatially coincident with the regions of the cloud that had column densities larger than A$_V$ = 30~mag (N$_{H_2}$$\sim$\replaced{3}{4.11}$\times$10$^{22}$~cm$^{-2}$).  These stars are most likely to be foreground stars because, even with deep exposures, it would be very difficult to detect \deleted{unreddened} stars in the NIR through such large extinctions.
				
\begin{figure} 
					\begin{center}
						\includegraphics[width = \columnwidth]{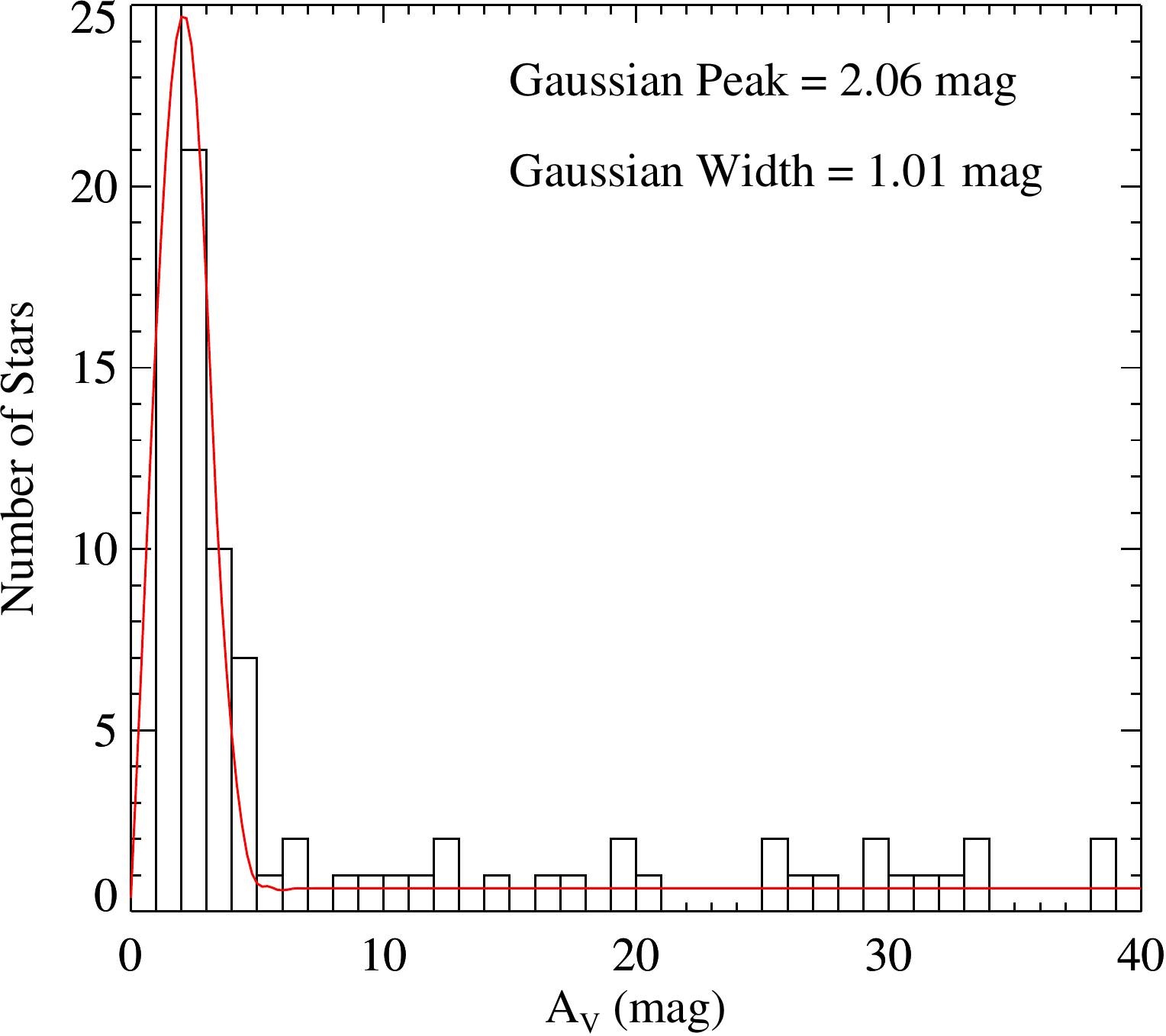}
						\caption{Distribution of estimated visual extinctions of polarization catalog stars projected within the region bounded by the A$_V$ = 30~mag contour of Figure \ref{figglimpsecold}.  A Gaussian plus a constant fit to the A$_V$ distribution, centered at 2~mag with a width of 1~mag, is overlaid in red.}
						\label{figforexred}
					\end{center}
				\end{figure}

				Figure \ref{figforexred} plots the distribution of stellar extinctions of stars \added{that are listed} in the Table \ref{table1} polarimetry catalog and are located within the region outlined by \added{the} A$_V$ = 30 mag \added{contour} in Figure \ref{figglimpsecold}.  These stars show predominantly low extinctions, with a small tail \replaced{of}{in the distribution extending to} larger extinctions.  The \added{binned} distribution of extinctions \added{(bin=1 mag)} was fit using a Gaussian plus a constant background.  The peak of the fit occurs at 2.06 mag with a Gaussian width ($\sigma$) of 1.01 mag.  \added{Fits made to different bin sizes resulted in similar peaks and widths.}  Based on the fit, we estimated that the foreground extinction is strongly bounded to be no more than 4.5 mag (\deleted{location of the} peak \added{mean extinction} plus $\sim$2.5$\sigma$).  \added{This value is a liberal estimate of the foreground extinction, and represents a conservative approach to assigning background stars with high confidence.}
				
				All stellar extinctions were estimated following the steps described in Section \ref{stellarextinctions}.  Based on their relative extinctions compared to the cloud extinctions at their locations, the stars were classified into three categories: `foreground,' `background,' and `unknown.'  Any star with extinction less than 4.5~mag was classified as `foreground.'  The extinctions of the remaining stars were reduced by 4.5~mag and compared to the cloud extinction at the stellar coordinates.  Stars with foreground-modified extinctions larger than the cloud extinction plus cloud extinction uncertainty (10\%) were classified as `background.'  Stars with modified extinctions less than the cloud extinction minus cloud extinction uncertainty were classified as `foreground.' Stars with extinctions that fell within the cloud extinction $\pm$ the cloud extinction uncertainty received distance classifications of `unknown.'  These stars were not included in further analysis because their relative locations with respect to the cloud could not be well-determined.
				
				The estimated stellar extinction (not \replaced{subtracted}{reduced} by 4.5 mag) of each star, along with the extinction estimation method used, and its distance assignment are listed in columns 9, 10, and 11, respectively, in Table \ref{table2}.  \added{The numbers of background, foreground, and unknown stars, with no polarization signal-to-noise cuts applied, were 10597, 6255, and 308, respectively.  The numbers of stars that exhibited high polarization SNR in at least one band were 900, 261, and 25 for background, foreground, and unknown stars, respectively.}
				
				Because foreground material is present along the line-of-sight to G28.23, the polarization measurements of the background stars needed to be corrected for any foreground polarization signal.  To remove this effect, we \added{first} assumed the foreground polarization was contained in a uniform layer.  \added{If more than one significant layer of material was present along the line of sight, the extinction values of the foreground stars would not be prominently peaked around one value.  Because this was the case toward the region of G28.23, it was safe to assume one uniform foreground layer.} 
				
				\replaced{and}{We} computed the variance-weighted average Stokes U$_E$ and Q$_E$ parameters of all foreground stars.  This foreground polarization signal was $\sim$0.5~\%.  The average foreground U$_E$ and Q$_E$ values ($\overline{U}$$_E$ = 0.40 \%, $\overline{Q}$$_E$ = -0.28~\%) were subtracted from the U$_E$ and Q$_E$ parameters of the individual background stars \added{prior to their use in subsequent analyses}.  The foreground, weighted U, Q uncertainties (of order $\sim$0.02~\%) were propagated into the foreground-subtracted U$_E$ and Q$_E$ uncertainties of the background stars.  We note that the polarization properties of the stars listed in Table \ref{table1} have not been foreground-modified, but can be \added{corrected} by following the above procedure.  The foreground-corrected polarization properties \added{of the background stars} are used in the rest of the analyses.

			\subsubsection{Properties of the Background Stars}
				The number of background stars with a high SNR in $K$-pol was 318, of which 90 were also detected as high SNR $H$-pol stars.  Within the entire 20$\times$20 arcmin FOV, 574 background stars were detected as high SNR $H$-pol stars, including those 90 stars.  Four high SNR $H$-pol stars in the 10$\times$10 arcmin FOV did not have corresponding $K$-pol entries.
				
\begin{figure*} 
					\begin{center}
						\includegraphics[width = 6in]{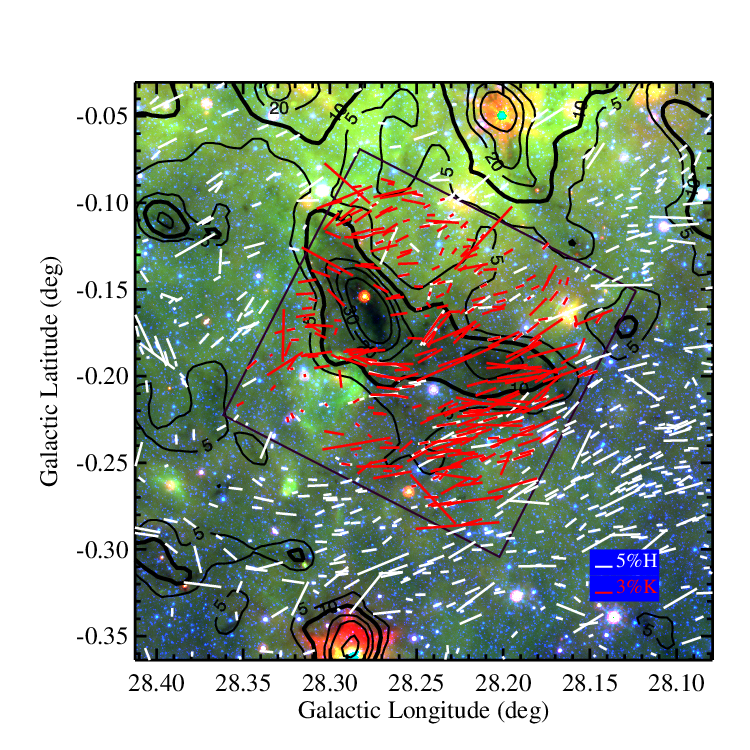}
						\caption{GLIMPSE+MIPSGAL 3-color background image, with column-density-derived A$_V$ contours overlaid in black.  The foreground-corrected polarizations of individual high SNR background stars are overlaid in red ($K$-pol) and white ($H$-pol).  The lengths of the vectors indicate the polarization percentage, and the scale of the vectors is shown in the bottom right.  The $H$-band polarization measurements trace the large-scale behavior of the plane-of-sky B-field, while the $K$-band polarizations probe the B-field of the intermediate \replaced{layers}{densities} of the cloud.  \added{ Ninety stars, which have both $H$ and $K$ high SNR polarization detections, have both white and red vectors shown.}}
						\label{figglimpseindvecbackforesub}
					\end{center}
				\end{figure*}

				Figure \ref{figglimpseindvecbackforesub} shows a GLIMPSE and MIPSGAL 3-color image of G28.23 overlaid with the \added{dust emission-based} column density-derived A$_V$ contours and the high SNR $H$ and $K$ stellar polarization vectors.  The polarizations probe to cloud A$_V$ values between 35 to 40 mag.  While the $K$-band polarizations cannot probe into the dense, inner \replaced{layers}{regions} of the IRDC, where A$_V$ values reach $\sim$100 mag, they probe to the intermediate \replaced{layers}{densities}.  Using a combination of $H$-pol and $K$-pol data, we trace the cloud plane-of-sky B-field orientation from the outer diffuse regions into the intermediate \added{extinction} regions of the cloud.  The large-scale polarization orientation across the FOV is preferentially parallel to the Galactic plane, although some \replaced{shift in this pattern is}{changes to this overall pattern are} seen near the IRDC.
				
				The following analysis and results use only the sample of $H$-pol and $K$-pol background stars.
				
		\subsection{Polarization Efficiency}
			Because the mechanism of spinning up dust grains to be oriented perpendicular to the intervening B-field relies on an anisotropic radiation field \citep{Lazarian2007}, one concern in using stellar polarimetry to probe B-fields in dense regions is whether the grains remain aligned with the field deep within the cloud \citep{Arce1998}.  
				
			\deleted{To find whether the accumulated polarization signal diminished as a function of extinction, we determined how polarization percentage behaved with increasing extinction.  Figure \ref{figcloudavvspolold} plots the relation between the stellar optical depth at $K$-band, $\tau_K$, and the polarization efficiency, PE, for high SNR $K$-pol stars, where PE is equal to the polarization percentage at $K$-band per $\tau_K$.  A$_V$ was assumed to equal $8.8~A_K$ (Majewski et al. 2011), and A$_K$ = 1.086~$\tau_K$ (Whittet 2003).}
				

			\deleted{The relation between A$_V$ and PE was best fit by a power law of index -0.4, with a linear correlation coefficient of -0.3.  Slopes shallower than -1 indicate that some polarization signal is being added as higher extinctions are probed.  This result, in answer to Question 1 posed in Section 1.2 implies that the grains remain aligned, at least to some degree, to the highest extinctions probed by the NIR in G28.23.}
			
\begin{figure*}  
					\begin{center}
						\includegraphics[width = 6in]{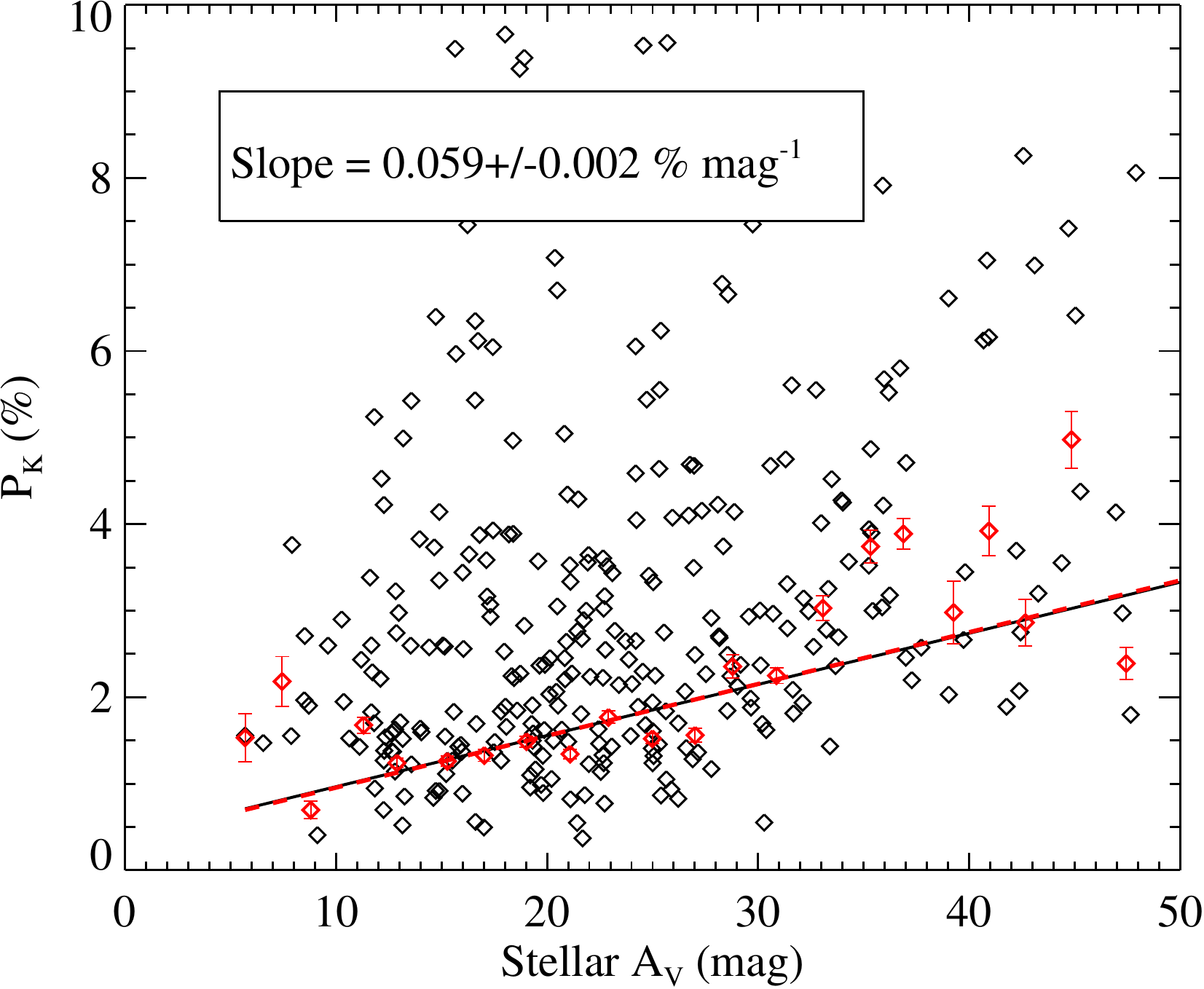}
						\caption{Stellar $K$-band polarization percentage  as a function of A$_V$ for the 318 high SNR background $K$-pol stars.  The slope of the linear relation between polarization percentage and A$_V$ is greater than zero, which indicates that the polarization percentage increases as a function of increasing extinction.  The individual stellar extinctions were separated into 2 mag wide bins, and the weighted average polarization percentage was found in each bin.  The binned averages and the best-fitting line between the binned average extinction and polarization percentage are overlaid in red.  The slope of this line is equal to 0.06.}
						\label{figcloudavvspol}
					\end{center}
				\end{figure*}

			\added{A relation between polarization percentage and extinction was found for the polarization stars in the G28.23 10$\times$10 arcmin FOV.  Figure \ref{figcloudavvspol} plots the stellar optical extinction, A$_V$, against the polarization percentage for high SNR $K$-pol stars.}	
						
				\added{The relation between A$_V$ and polarization was fit by a line, variance-weighted by the polarization percentage uncertainties.  The best-fitting line had a slope of 0.059$\pm$0.002 \% mag$^{-1}$.  The individual polarization percentages versus stellar extinctions show significant scatter.  Therefore, the polarization percentages of the individual stars were binned into extinction bins of width = 2 mag, and the variance-weighted average of the stellar polarization percentages in each bin were computed.  These binned polarization percentages and the best-fitting line are overlaid in Figure \ref{figcloudavvspol} in red.  The slope of the best-fitting line to the binned points is 0.060 $\pm$0.002 \% mag$^{-1}$, which agrees with the slope of the relationship of the individual points.  
					
				This positive slope between the P$_K$--A$_V$ relation is less than the slope derived by \citet{Arce1998} for the polarization measurements of stars in the ISM near the Taurus molecular cloud.  Using polarimetric observations taken at 766 nm, they found that P$_{766nm}$$\sim$3.58$E_{B-V}$, assuming A$_V$ = 3.1$E_{B-V}$, for A$_V$ less than 1.3 mag.  Accounting for the expected difference in polarization percentage between the two wavelengths of 766 nm and $K$-band at 2.2 $\mu$m following the Serkowski Law, and using an R$_V$ equal to 5.5, as was done in this study, their relation would be equivalent to P$_K$$\sim$0.12~A$_V$.  This slope is about a factor of two larger than the one derived in this study.  The difference between the two slopes may be due to a decrease in polarization efficiency at the large opacities of G28.23, which is embedded in $\sim$2 mag of visual extinction in the Galactic plane.  Taurus, on the other hand, is located off the plane, and exposed to the interstellar radiation.}  
				
				\added{Slopes larger than zero indicate that some polarization signal is being added as higher extinctions are probed.  The results obtained here, in answer to Question 1 posed in Section \ref{methodology}, imply that the grains remain aligned, at least to some degree, to the highest extinctions probed by the NIR in G28.23.}

		
		\subsection{UQ Averaging}
			Using only high SNR stars ignores the stars with lower signal-to-noise ratios.  While these stars are not significant individually, they can be averaged to boost average polarization signal-to-noise \citep{Clemens2012c}.  Therefore, we spatially averaged the polarimetric information of all of the background stars, \added{separately for the $K$ and $H$-pol stars,} to create \deleted{a} smoothed \added{$H$ and $K$} polarization \replaced{map}{maps} of the cloud.  The \replaced{map was}{maps were} gridded into pixels of 30$\times$30 arcsec, and all stars classified as background were used.  The variance-weighted average U$_E$ and Q$_E$ Stokes parameters of the stars located within each pixel were computed \added{(separately for the $H$ and $K$-pol measurements)}, from which debiased (Section \ref{catalog of nir stars}) average polarization \added{percentages} and Equatorial PAs were estimated.  The polarizations computed for each pixel were therefore independent.  These Equatorial PAs were rotated into GPAs.
			
\begin{figure*}  
				\begin{center}
					\includegraphics[width = 6in]{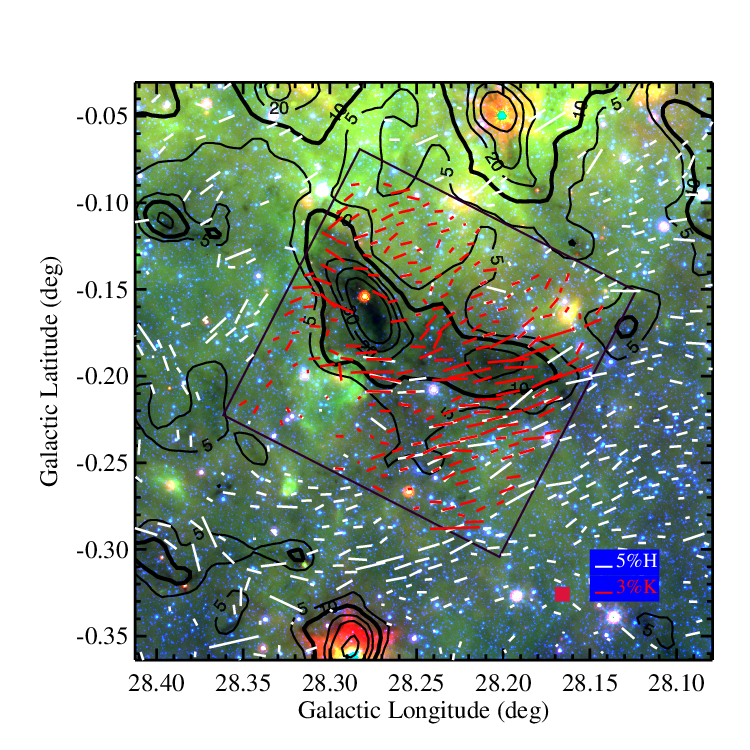}
					\caption{Similar to Figure \ref{figglimpseindvecbackforesub}, but with the spatially averaged Stokes U$_E$ and Q$_E$ used to compute the high SNR polarization vectors shown.  The U$_E$ and Q$_E$ parameters of the individual stars were averaged over 30$\times$30~arcsec bins \added{for each of the two wavebands}.  This bin size is represented by the red box in the bottom right.}
					\label{figglimpseuqavgbackforesub}
				\end{center}
			\end{figure*}

			Figure \ref{figglimpseuqavgbackforesub} is similar to Figure \ref{figglimpseindvecbackforesub}, but with the spatially-averaged polarization vectors plotted.  Only the polarization values of pixels where the polarizations were equal to or greater than 2.5 times their \replaced{associated}{propagated} uncertainties are shown.  In this average polarization map, \added{random components of the polarizations are averaged over, and} the large-scale patterns of the polarizations become more evident than in Figure \ref{figglimpseindvecbackforesub}.  The polarization GPAs are preferentially perpendicular to the \deleted{main} cloud major axis farther from the cloud and, especially at larger longitudes, \added{the polarization orientations} twist to become more parallel \replaced{as they approach}{to} the cloud \added{major axis as vectors that are closer to the cloud spine are considered}.
		
		\subsection{Relative GPA Orientations}\label{relativegpaorientations}
			To explore the PA orientation patterns further, we separated the polarization map into four cloud-centered quadrants and compared the GPA distributions of the individual background stars \added{located} in each of these regions.
			
\begin{figure*}  
				\begin{center}
					\includegraphics[width = 6in]{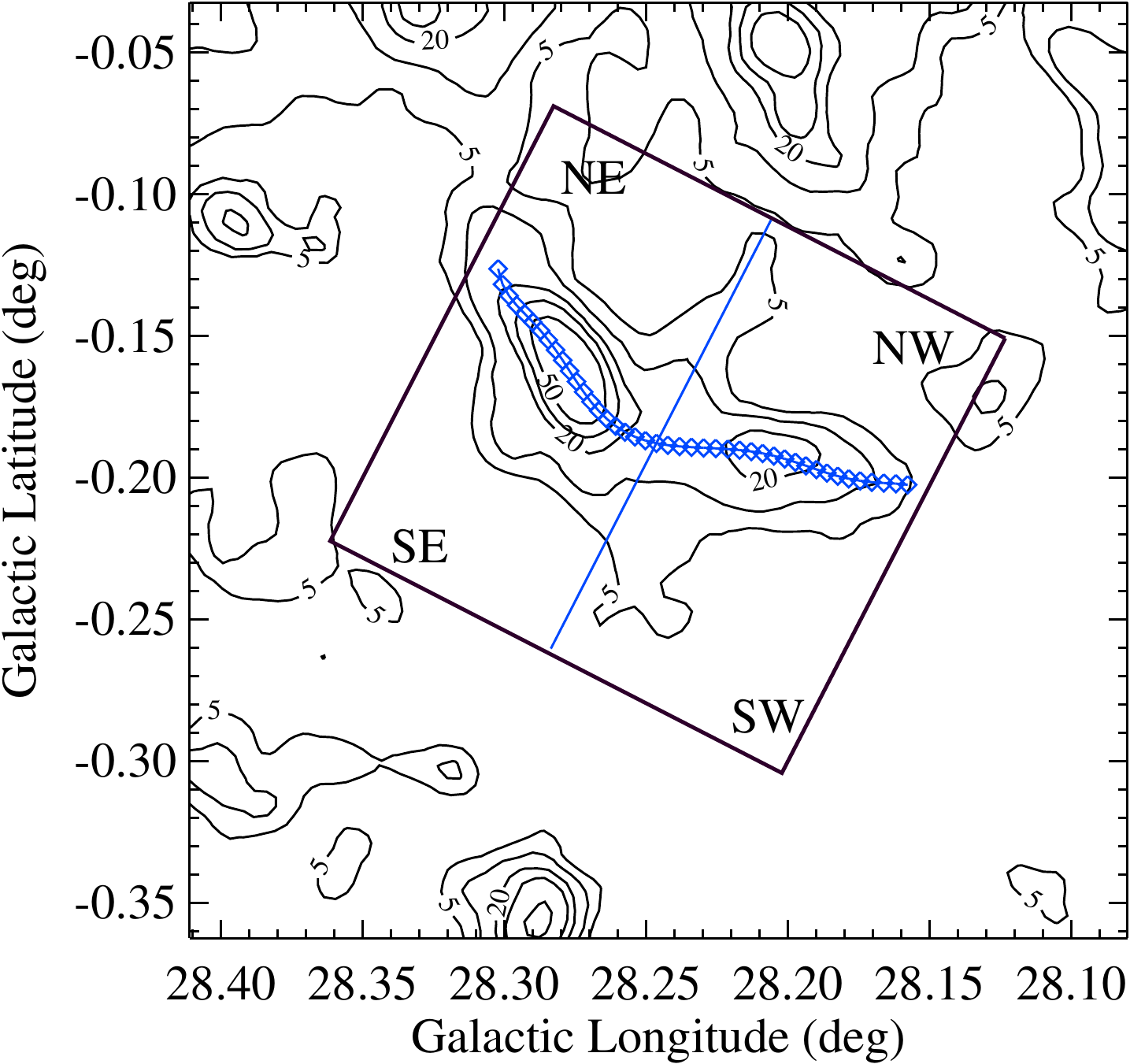}
					\caption{Column-density-derived A$_V$ contours of G28.23 are shown in black, with the box representing the 10$\times$10 arcmin K-pol FOV overlaid.  The spine of the cloud is overlaid as blue diamonds, which separates the cloud into Northern and Southern components.  The cloud is further separated into East and West components by a blue line, of constant R.A., running through the spine midpoint.}
					\label{figpaslicelinespine}
				\end{center}
			\end{figure*}

			An F-test supported 7th order polynomial was fit to the column density map to define a cloud `spine' along its long axis \citep[e.g.,][]{Marchwinski2012, Cashman2014}.  The polynomial was fit to the pixel locations of the peak values of the column density map along the \replaced{decl.}{declination} axis, so the spine points were spaced $\sim$37~arcsec apart.  Figure \ref{figpaslicelinespine} shows the cloud column density-derived A$_V$ contours, with this spine overlaid.  The spine splits the cloud into `Northern' and `Southern' (Galactic) components.  In addition, we also separated the cloud into `Eastern' and `Western' sections by \added{using} a line running parallel to the R.A. axis through the spine midpoint.  This line separates the northern dense region of the cloud, which hosts several massive starless cores \citep{Sanhueza2013} from the less dense, southern elongated extension \added{of the cloud}.
			
\begin{figure*}  
				\begin{center}
					\includegraphics[width = 6in]{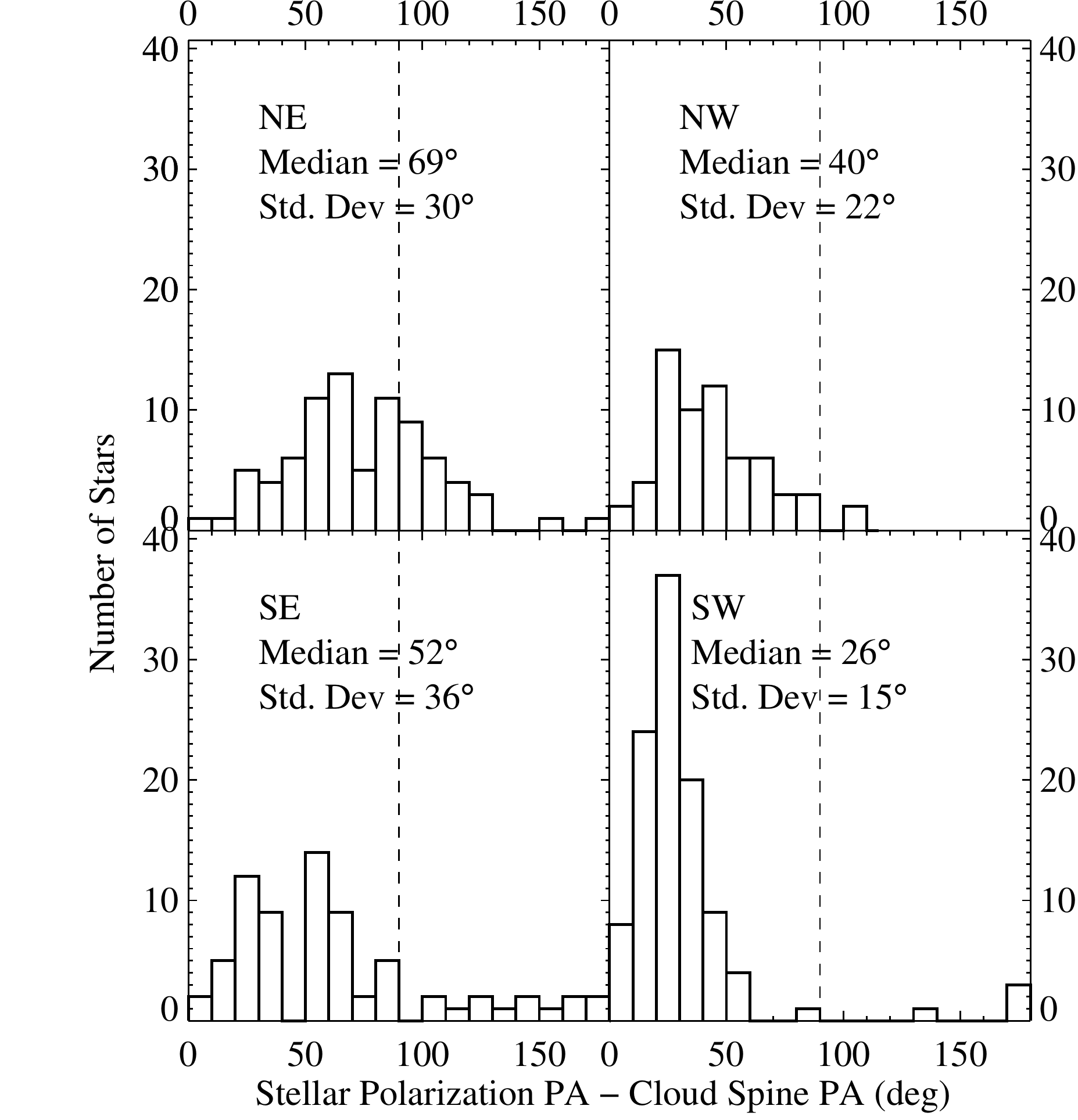}
					\caption{Distributions of polarization PAs of individual high SNR stars relative to the orientation of the spine of G28.23.  The four distributions correspond to the quadrants labeled in Figure \ref{figpaslicelinespine}.  The median and standard deviation of each distribution are listed.  The dashed line in each panel represents 90$\degr$, the perpendicular to the cloud orientation.}
					\label{figdelpasn25}
				\end{center}
			\end{figure*}

			\replaced{The PA difference of}{For} each high SNR $K$-pol and $H$-pol star in the 10$\times$10 arcmin FOV, \added{the PA} was found relative to the orientation of the cloud spine at the closest point to the star.  \added{The $K$ and $H$-pol PAs were treated as one sample.  For stars with both $K$ and $H$-pol measurements, their $K$-band PAs were used, and for stars with a high SNR detection in just one band, the PA of the polarization measurement in that band was used.  This method of selecting stellar polarimetric information is hereafter referred to as `$K$-leading'.}  The PA differences ranged from 0 to 180$\degr$, where 90$\degr$ signifies PAs that are perpendicular to the cloud spine, and 0 or 180$\degr$ signify PAs that are parallel to the spine.  \added{While the difference between two angles with no preferential orientation cannot exceed 90$\degr$, a preferential orientation was assigned for the relative PAs between the cloud spine and polarization measurements.  This assumption resulted in PA differences between 0 and 180$\degr$ instead of 0 and 90$\degr$.  This assumption was made because angles measured between 0 and 180$\degr$ could reveal differences in the relative PA distributions among the four quadrants.}  Figure \ref{figdelpasn25} shows the distributions of relative PA orientations for stars in the four quadrants.  
			
			The four distributions show different \deleted{preferential} relative PA patterns.  \added{Standard deviations were calculated for the distributions.  Because PA distributions are directional or circular, and wrap around 0 and 180$\degr$ (a PA of 190$\degr$ is equivalent to 10$\degr$), the standard deviations of the distributions shown would not necessarily represent the true deviations in the data.}  The standard deviation of each distribution was found by shifting the whole distribution by increments of 10$\degr$ and wrapping PAs greater than 180$\degr$ \deleted{(where a PA of 190$\degr$ is equivalent to a PA of 10$\degr$)}, finding the standard deviation of each shifted distribution, and selecting the \replaced{lowest}{least} deviation.  \added{Increments other than 10$\degr$ were also tested, but resulted in very similar deviation estimates.}  The Northeast (larger $l$ and $b$ values) distribution peaks around a relative orientation of 69$\degr$ (\added{i.e.,} more perpendicular) with a standard deviation of 30$\degr$.  The Southwest distribution peaks around 26$\degr$ (more parallel), with a standard deviation of 15$\degr$.  The Northwest and Southeast distributions are not strongly skewed \replaced{in}{toward} either \replaced{direction}{parallel or perpendicular}, though the majority of stars in both distributions have relative PAs between 0 and 90$\degr$.  The Southeast distribution is somewhat more preferentially perpendicular, though it has a larger standard deviation of 36$\degr$.  \added{Interestingly, there is a large difference in the preferred relative orientations between the Northeast and Northwest quadrants, as well as between the Southeast and Southwest quadrants.  The median relative PA of each pair (Northeast to Northwest and Southeast to Southwest) of quadrants differs by nearly 30$\degr$.}
		
			\subsection{PA Dispersion}
				\replaced{To estimate the degree of order in the polarization measurements, as}{As} a partial proxy for the plane-of-sky B-field strength under the CF method, we calculated the PA dispersion across the cloud.  \replaced{using}{This dispersion calculation used} the 322 high SNR $K$ and $H$ polarizations in the 10$\times$10 arcmin FOV \added{(selecting the polarization information of the stars via the K-leading procedure)}.
				
				A large-scale PA pattern was removed from the polarization \replaced{map}{PAs} prior to calculating the PA dispersions, as the estimated dispersion of the B-field should only include the turbulent motions of the gas \citep[e.g.,][]{Ostriker2001}.  To create the large-scale pattern, the individual stellar PAs \added{($K$-leading)} were smoothed using variance weighting.  \replaced{, with}{Bin} centers \added{were} separated by 45 arcsec \deleted{and Gaussian kernels}.  \added{Smoothing with a Gaussian kernel of} \deleted{with} $\sigma$ equal to $\sim$38 arcsec \added{was used} to achieve Nyquist sampling.  A range of center separations and corresponding Nyquist-sampled Gaussian kernel sizes were tested, but the results did not vary significantly.  \replaced{This}{The resulting} smoothed PA map \deleted{, with 45 arcsec bin separations,} was then interpolated to the size of the Mimir instrument platescale, 0.58 arcsec per pixel.  The smoothed and interpolated PA \replaced{pattern was}{map values were} then subtracted from the individual stellar GPAs \added{at the position closest to each star}.
				
				The PA dispersions of \replaced{the large-scale, pattern-subtracted stars}{these large-scale corrected GPAs} were calculated using large, overlapping 120$\times$120 arcsec bins with center separations of 60 arcsec.  Such large bins were necessary to ensure that enough stars with high signal-to-noise were used in the dispersion calculations.  Dispersions (\added{unweighted} standard deviations of the PA distributions) and propagated uncertainties were calculated only for bins with at least seven high SNR stars.  \added{The PA distributions were dealiased, as described in Section \ref{relativegpaorientations}, and the standard deviation was found for each iteration.  The dispersion was set equal to the minimum standard deviation of the 18 iterations.  The uncertainty was the propagated uncertainty of the standard deviation.}
				
				Following \citet{Hildebrand2009}, the uncertainties of the PA dispersions were subtracted in quadrature from the dispersions.  This step was performed to correct the dispersions for the bias added by observational uncertainties, for which the median uncertainty was $\sim$3$\degr$.  The median post-correction signal-to-noise of the PA dispersions was seven.  \added{The PA dispersions and their uncertainties are listed in Table \ref{table3}.  The median PA dispersion was 20$\degr$, with a standard deviation of 7$\degr$.}
				
\begin{figure*}  
					\begin{center}
						\includegraphics[width = 6in]{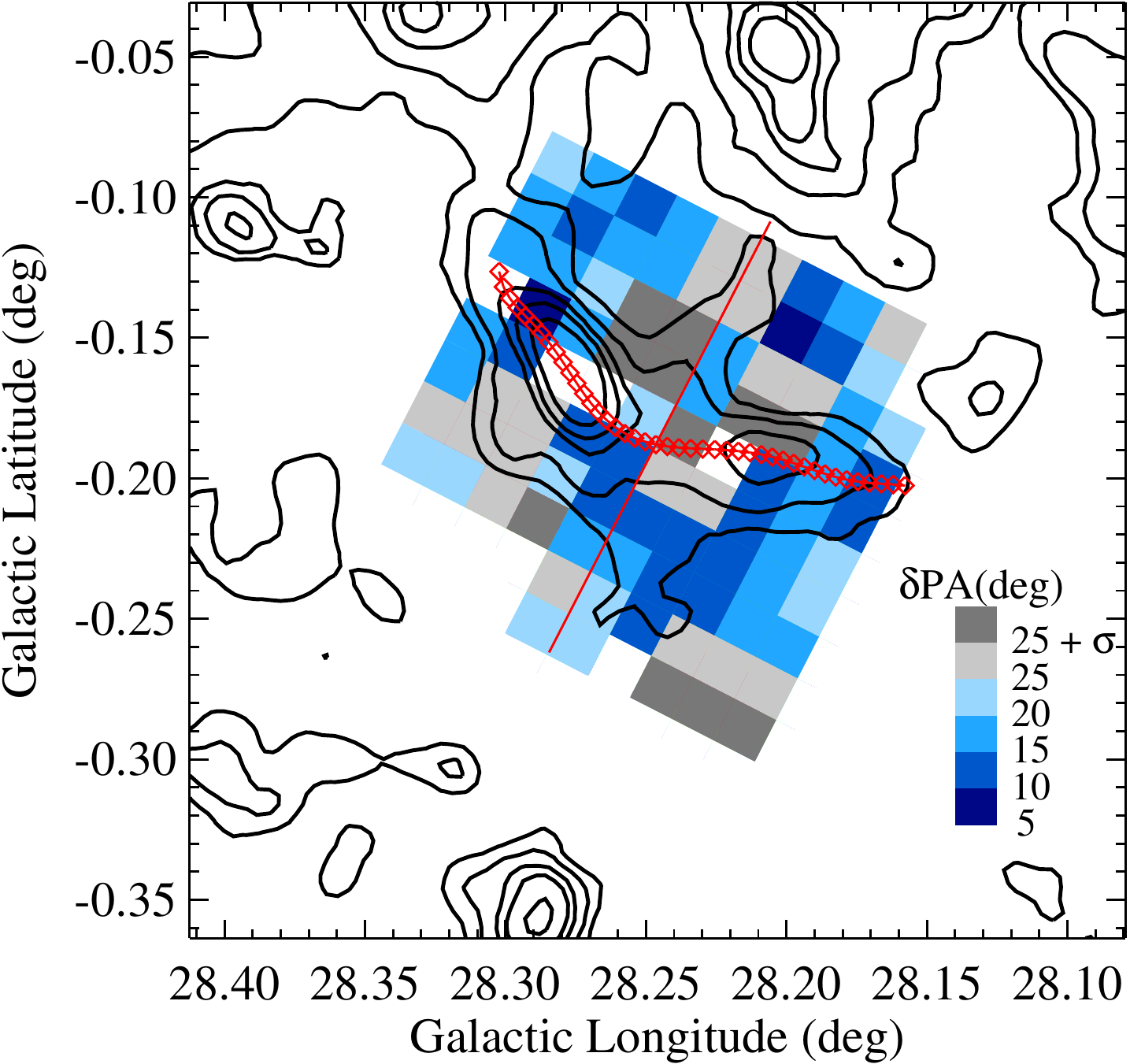}
						\caption{PA dispersion map of G28.23, using only high SNR \added{$H$ and $K$-band} stars.  \added{For stars with both $H$ and $K$-band high SNR polarization detections, their $K$-band GPAs were used in the dispersion calculations.}  The blue filled pixels indicate regions where the dispersions are less than 25$\degr$, with darker hued-colors corresponding to lower dispersions.  Regions of light gray correspond to dispersions that are larger than 25$\degr$, but the dispersion values minus their uncertainties are less than or equal to 25$\degr$.  Dark gray regions indicate where the dispersions minus their uncertainties are larger than 25$\degr$, while white regions correspond to locations where the numbers of stars were not sufficient to calculate a dispersion.  Column-density derived A$_V$ contours are overlaid in black, and the cloud spine and East-West dividing line are shown in red.}
						\label{figpadisp}
					\end{center}
				\end{figure*}

				\added{The plane-of-sky B-field strengths were calculated using the CF method (described below) in the pixels where the PA dispersions, or the PA dispersions minus their uncertainties, were below 25$\degr$, as recommended by \citet{Ostriker2001}.}  The corrected PA dispersion map is shown in Figure \ref{figpadisp}.  \deleted{The median PA dispersion was 20$\degr$, with a standard deviation of 7$\degr$.}  The blue regions show where PA dispersions were below 25$\degr$.  The light gray regions indicate bins where the PA dispersions themselves were above 25$\degr$, but the PA dispersions minus their uncertainties were less than or equal to 25$\degr$.  These blue and light gray regions are where the plane-of-sky B-field strengths were calculated using the CF method \deleted{(Ostriker et al. 2001)}.  The dispersions could not be calculated in the dense center regions of the cloud due to the lack of stars, \replaced{and were not calculated}{which are shown in white.  Regions} where the PA dispersions minus their uncertainties were \deleted{still} above 25$\degr$ \deleted{.  These zones} are shown as \deleted{white and} dark gray pixels \deleted{, respectively}.
				
			\subsection{Cloud Volume Density}\label{cloud volume density}
				\added{The spatial distribution of cloud volume density is necessary to estimate the plane-of-sky B-field strength using the CF method.  The average volume density map of the cloud was derived using the map of dust emission-based column density.  We assumed that the cloud was oriented with its long axis in the plane of the sky.}
				
\begin{figure*}  
					\begin{center}
						\includegraphics[width = 6in]{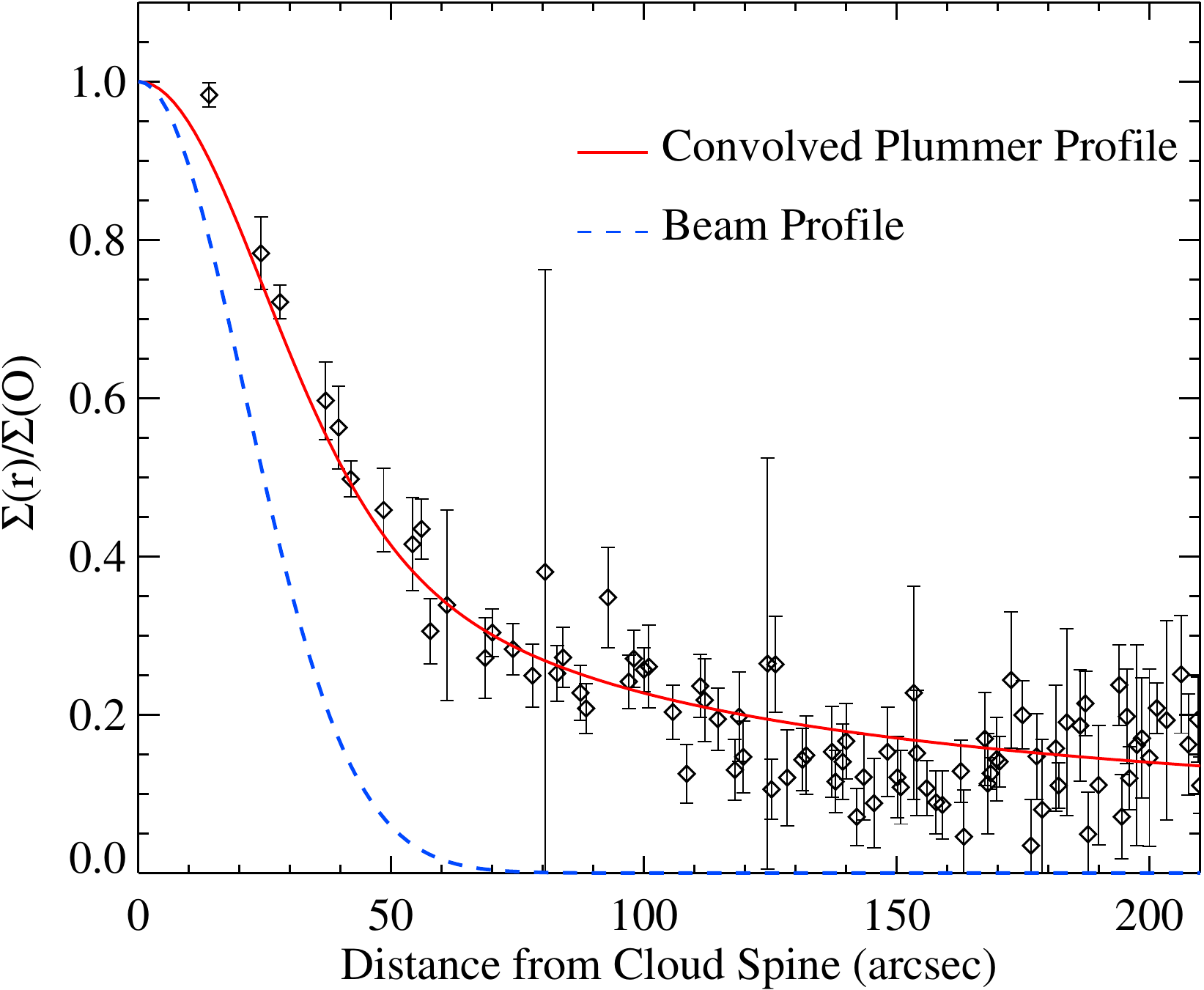}
						\caption{\replaced{Plummer-like profile fit to the normalized column density points in each pixel of the column density map.  The distance is calculated from the closest point along the cloud spine.}{Normalized column density profile of G28.23.}  Each black diamond represents the median \added{normalized} value \replaced{in bins along the distance, and the}{of the pixels falling into each distance bin in the column density map, where the distance of each pixel is calculated to the closest point on the cloud spine.  The} error bars enclose $\pm$1 standard deviation in each \added{distance} bin.  \added{A Plummer-like profile was fit to the normalized column density points.}  The Plummer fit is shown as the solid red line, and the Herschel 500~$\mu$m beam profile is shown by the dashed blue line.  \added{The profile of the column density of the cloud is wider than the {\it Herschel} 500$\mu$m beam profile.}}
						\label{coldnormplumfit}
					\end{center}
				\end{figure*}

				\deleted{We created an average volume density map of G28.23 by first generating a 3-dimensional volume density model of the cloud.  A Plummer-like model was fit to the normalized cloud column density ($\Sigma(r)/\Sigma(0)$) profile shown in Figure \ref{coldnormplumfit} (e.g., Equation 1 of Arzoumanian et al. 2011)}
				
				\added{Before calculating an average volume density map of G28.23, a 3-dimensional (3D) volume density model of the cloud was first created.  A normalized column density profile ($\Sigma(r)/\Sigma(0)$), shown in Figure \ref{coldnormplumfit}, was created using the cloud column density values.  The cloud spine was assumed to be located in the plane of the sky (one fixed distance in the line of sight), and the density profile was assumed to be cylindrically symmetric about the cloud spine.}  
				
				A Plummer-like model was fit to the normalized column density profile, following Equation 1 of \citet{Arzoumanian2011}:
				\begin{align}
				\rho({\bf r}, k) = \frac{\rho_c}{[1 + ({\bf r}/R_{flat})^{2}]^{\frac{p}{2}}},\\
				\Sigma(r) = C_p~\frac{\rho_{ck} R_{flat}}{[1 + (r/R_{flat})^{2}]^{\frac{p-1}{2}}},\\
				C_p = \int_{-\infty}^{\infty} \frac{(1/R_{flat})dr}{[1+(r/R_{flat})^2]^{\frac{p}{2}}},
				\end{align}
				\replaced{where $\rho_c$ is the central gas volume density of the cloud, k is the distance along the cloud spine, and $p$ is the profile index.  R$_{flat}$ represents the radius of the central flat portion of the column density profile, and $r$ is the projected distance to the closest location, k, on the cloud spine.  $\Sigma(r)$ is the mass surface density, which is equal to $\mu$m$_H$$N_{H2}$.  The volume density, which is dependent on the distance from the cloud spine, is $\rho_{p}(r, k)$.  The central density was allowed to vary with k.  The best R$_{flat}$ and $p$, fit for all the pixels in the column density map, were 0.8$\pm$0.3 pixels ($\sim$30 arcsec) and 2.0$^{+0.1}_{-0.3}$, respectively.  The central densities along the cloud spine ranged from 4.2$\times$10$^{3}$ to 3.10$\times$10$^{4}$ H$_2$ cm$^{-3}$, with a median uncertainty of 200 H$_2$ cm$^{-3}$.}
				{where $k$ is the distance along the cloud spine, ${\bf r}$ is the 3D distance to the cloud spine, $r$ is the projected distance to the closest location, $k$, to the cloud spine, $\rho_{ck}$ is the central gas volume density of the cloud,  and $p$ is the profile index.  R$_{flat}$ represents the radius of the central flat portion of the column density profile.  $\Sigma(r)$ is the mass surface density, equal to $\mu$m$_H$$N_{H2}$.  The volume density, which depends on the distance from the cloud spine, is $\rho_{p}({\bf r}, k)$.  The central density was allowed to vary with $k$.  It was constrained by the fit to the column density profile, and was fit point by point along the cloud spine to reproduce the column density values.  The best R$_{flat}$ and $p$, fit using all of the pixels in the column density map (within the central 10x10 arcmin box in Figure \ref{figglimpsecold}), were found to be 0.8$\pm$0.3 pixels ($\sim$30 arcsec) and 2.0$^{+0.1}_{-0.3}$, respectively.  In comparison, the profile index would be equal to 4 for an isothermal cylinder in equilibrium \citep{Ostriker1964}, indicating that there may be some magnetic support for G28.23 \citep{Fiege2000}.  The central densities along the cloud spine were found to range from 4.2$\times$10$^{3}$ to 3.1$\times$10$^{4}$ H$_2$ cm$^{-3}$, with uncertainties of about 200 H$_2$ cm$^{-3}$.}
			
				This method returned a 3D \added{model} volume density data cube, with voxel \deleted{cube} side sizes equal to the column density pixel size (\deleted{each side = }37~arcsec $\sim$ 0.9~pc at a distance of 5.1~kpc).  \deleted{The cloud spine was located at one fixed distance along the line of sight (a Z axis), and the density profile was assumed to be cylindrically symmetric about the cloud spine.  In other words, for simplicity, the spine of the cloud was assumed to be contained within the plane of the sky.}
				
				To calculate an average volume density map, a \added{limiting} cloud boundary was needed, since the Plummer profile of the cloud extended to infinity.  \replaced{The cloud boundary was assumed to be where the maximum volume density along the light of sight just reached 50~H$_2$~cm$^{-3}$.  This value was chosen because it was the median volume density value in the plane of the cloud spine that was spatially coincident with the A$_V$ contour of 2~mag.  G28.23 is not an isolated cloud, but instead is embedded in diffuse material, which was estimated to have an extinction of 2~mag.  This estimate is foreground-corrected, as the dust column density maps were corrected for such effects.  This extinction of 2~mag was estimated as the extinction where the radial cloud column density profile plateaued.}{The cloud boundary was assumed to be located where the column density radial profile reached a plateau.  This plateau occurs at an estimated extinction level of A$_V$$\sim$2~mag, indicative of the fact that G28.23 is not an isolated cloud, but is instead embedded in diffuse material.  Using the 3D density model, we determined that the A$_V$ contour of 2 mag corresponds to a median volume density of 50~H$_2$~cm$^{-3}$ in the plane of the sky.  Based on this correspondence, the cloud boundary was assumed to be where the maximum volume density along the line of sight just reached 50~H$_2$~cm$^{-3}$.}
				
				\deleted{The volume density along the line of sight was averaged for voxels that had density values greater than 50~H$_2$~cm$^{-3}$.}  \added{To reach an A$_V$ of 2 mag with an average volume density 50~H$_2$~cm$^{-3}$, a column of $\sim$20 pc is needed, which is comparable to the size of a giant molecular cloud (GMC), and is perhaps indicative of the GMC in which G28.23 is embedded.}  This \added{50~H$_2$~cm$^{-3}$} boundary is more liberal than ones used in previous studies, which used the cloud FWHM to estimate the cloud depth along the line of sight \citep[e.g.,][]{Crutcher2004}.  For G28.23, \added{the FWHM was found by fitting Gaussian profiles to column density slices across the cloud minor axis.  The} cloud column density FWHM corresponds spatially to a median volume density in the plane of the cloud spine of $\sim$550~H$_2$~cm$^{-3}$.  \replaced{The FWHM was found by fitting Gaussian profiles to column density slices across the cloud minor axis.}{Using the FWHM boundary in calculating the average volume density map might be overly restrictive in the case of G28.23.}  The selected boundary of 50~H$_2$~cm$^{-3}$ corresponds to a cloud width of $\sim$4$\sigma$.  \added{The effects of using different boundaries are explored further in Section \ref{systematicuncertainties}.}
				
\begin{figure*}  
					\begin{center}
						\includegraphics[width = 6in]{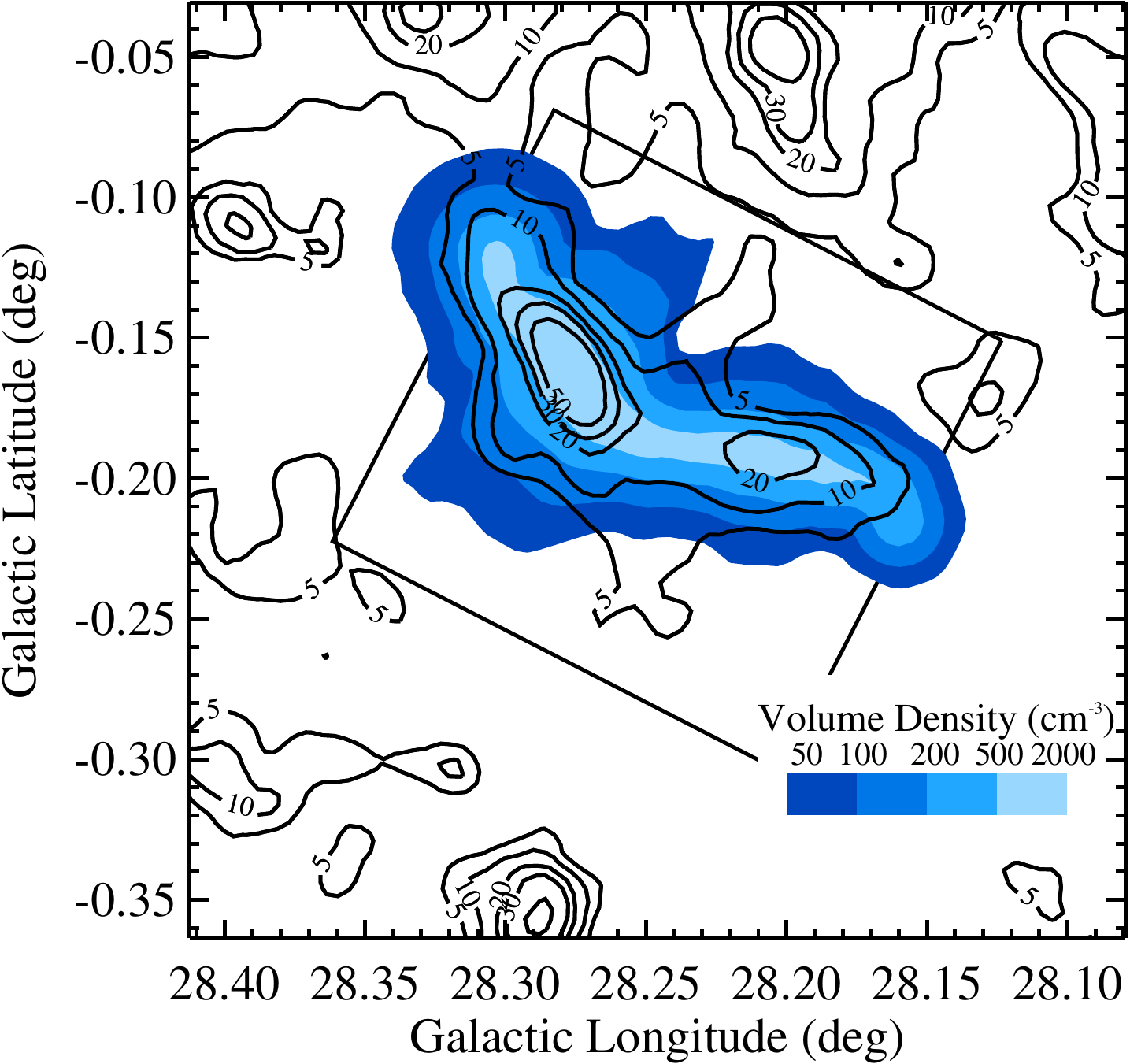}
						\caption{Average volume density map of G28.23, with column-density-derived A$_V$ contours overlaid in black.  The filled blue contours represent the values of the average volume density, with darker blue hues indicating lower volume density.}
						\label{figavgvoldenmap}
					\end{center}
				\end{figure*}

				\added{The volume density along each line of sight was averaged for voxels that exhibited density values greater than 50~H$_2$~cm$^{-3}$.}  The average H$_2$ volume density map of G28.23 is shown in Figure \ref{figavgvoldenmap} \added{as concentric filled blue contours}.  \replaced{The boundary of the density map}{The outermost density boundary} indicates where the average volume density just reaches 50~H$_2$~cm$^{-3}$.  The maximum average volume density is 1.9$\times$10$^{3}$~H$_2$~cm$^{-3}$.  If the boundary of 550~H$_2$~cm$^{-3}$ corresponding to the FWHM of the column density were used instead, the maximum average volume density becomes 6.1$\times$10$^{3}$~H$_2$~cm$^{-3}$.  The larger boundary \added{corresponding to the cutoff} of 50~H$_2$~cm$^{-3}$ allows the B-field \replaced{of}{across} a larger area \added{of the cloud (factor of $\sim$3)} to be computed.

		\subsection{Plane-of-Sky B-field Strength}	
			To estimate the plane-of-sky B-field strength (B$_{POS}$), we used the CF method, modified as recommended by \citet{Ostriker2001}:
			
			\begin{equation}\label{eqnbpos}
			B_{POS} = f~\sqrt{4\pi\rho}~\frac{\sigma_v}{\sigma_{\phi}},
			\end{equation}
			where $\rho$ is the mass density \deleted{of the cloud} (in grams cm$^{-3}$), $\sigma_v$ is the one-dimensional non-thermal gas velocity dispersion (in cm~s$^{-1}$), $\sigma_{\phi}$ is the polarization PA dispersion (in radians), $f$ is a correction factor, and $B_{POS}$ is the plane-of-sky B-field strength in units of Gauss.  Following \citet{Ostriker2001}, $f$ is closest to 0.5, and the method is only valid along directions where the PA dispersion is less than 25$\degr$ (0.44 radians).  
		

			The gas velocity dispersion\added{s} \replaced{was}{were} estimated \replaced{from}{by evaluating} the GRS $^{13}$CO spectral \added{line} data cubes, which have an angular resolution of 46 arcsec \added{(with pixel separations of 23 arcsec)} and spectral resolution of 0.2 km s$^{-1}$ \citep{Jackson2006}. 
			
			\added{The map of the $^{13}$CO integrated intensity was overlaid onto the dust-emission-based column density map to determine whether the $^{13}$CO traced the dust column density.  The integrated intensity of each pixel in the spectral data cube was calculated by summing the total intensity over $\pm$10 km s$^{-1}$ of the peak radial velocity of G28.23, $\sim$80 km s$^{-1}$.  Figure \ref{figcoiicoldmap} shows the $^{13}$CO integrated intensity with column density-derived A$_V$ contours overlaid.  The $^{13}$CO emission at 80 km s$^{-1}$ traces the same region as the dust emission-based column density along the line of sight, indicating that the $^{13}$CO spectra can be used to estimate the gas velocity dispersion of the cloud.
				
				\begin{figure*} 
					\includegraphics[width=6in]{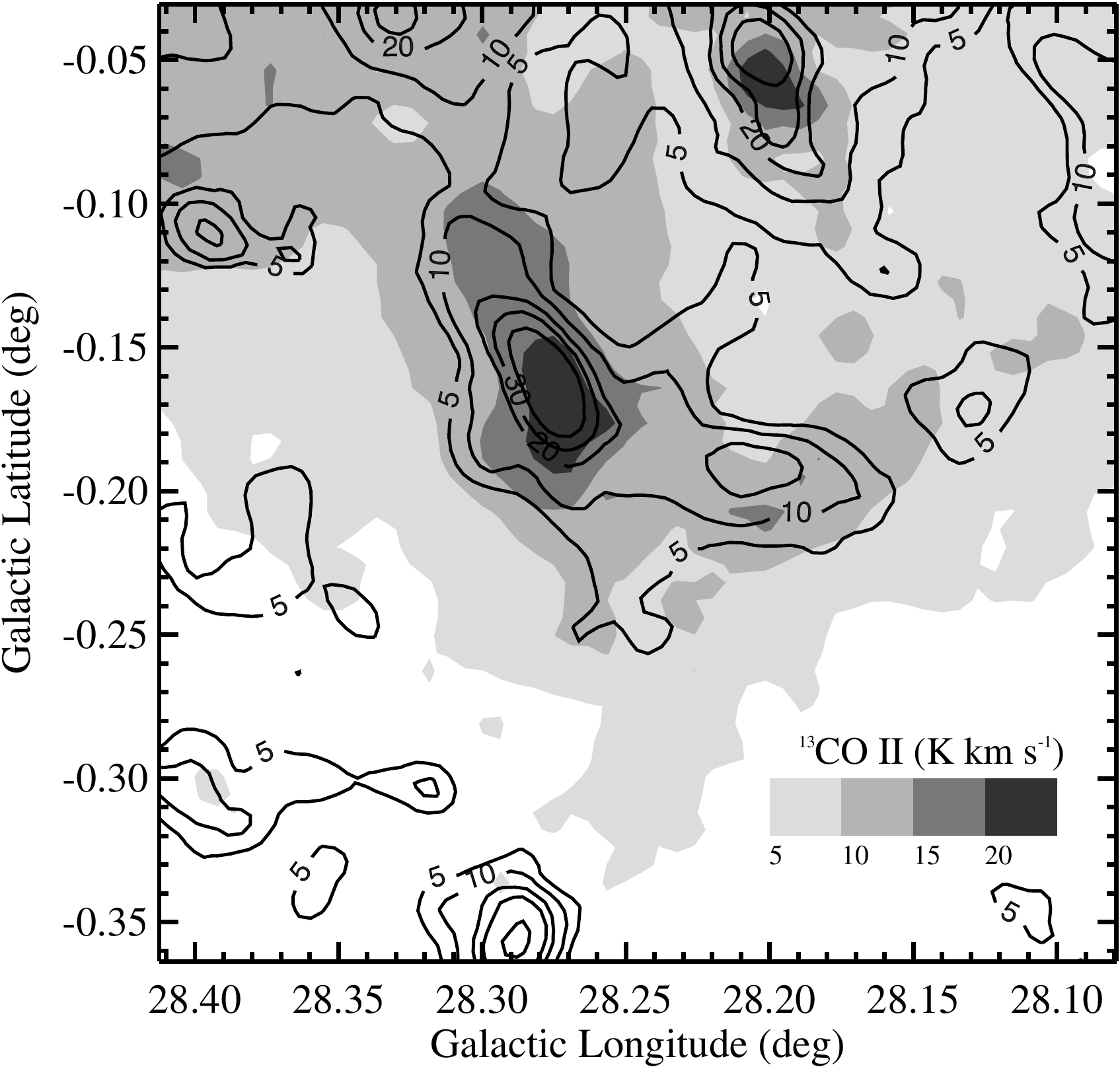}
					\caption{$^{13}$CO integrated intensity \added{gray-scale, filled} contours with dust-emission based column density-derived A$_V$ contours overlaid.  \replaced{The integrated intensity is shown as the grey-filled contours, whose values correspond to the legend at the}{Integrated intensity legend is at bottom right.}}
					\label{figcoiicoldmap}
				\end{figure*}
				}
			 \replaced{The spectral line seen in each pixel, across the 10$\times$10 arcmin FOV, was fit with a single Gaussian component, with the Gaussian sigma representing the velocity dispersion, since thermal contributions are negligible for these wide lines.}{The v$_{lsr}$ of G28.23 peaks at about 81 km s$^{-1}$ \citep{Sanhueza2013}.  A secondary smaller peak at 73-34 km s$^{-1}$ can be detected in the $^{13}$CO spectra in some regions near the cloud.  The 81~km~s$^{-1}$ line feature was fit with a Gaussian profile for each GRS spatial pixel falling within the 10$\times$10 arcmin FOV.  Where detected, the secondary component at 73 km~s$^{-1}$ was also fit with a Gaussian profile.}  
			
			\added{Only fit information of the 81 km s$^{-1}$ Gaussian-fitted feature was used in the CF method.}  The median of the velocity dispersions \added{of the GRS pixels, found from the Gaussian fits,} was $\sim$3 km s$^{-1}$, and the median SNR of the velocity dispersions was 12.  \added{The Gaussian sigma fit was interpreted to represent the velocity dispersion, since thermal contributions are negligible for these wide lines.}
				
			\added{In regions with high enough densities, $^{13}$CO could become optically thick, in which case the line would become saturated and the line width would exceed the gas velocity dispersion.  However, the B-field could only be calculated in regions with enough background stellar polarization probes to calculate PA dispersions, which excluded the \replaced{high}{highest} density regions of G28.23.  Examination of the $^{13}$CO spectra in the regions for which PA dispersions were calculated indicated that the lines were not self-absorbed.}
			
			\replaced{The angular resolution of the B$_{POS}$ map was limited by the resolution of the PA dispersion map shown in Figure~\ref{figpadisp}.  Both the velocity dispersion and volume density maps were resampled to the resolution of the PA dispersion map, with bins of 60$\times$60 arcsec.}{The angular resolutions of the $^{13}$CO velocity dispersion, average volume density, and PA dispersion maps were 46 arcsec, 37 arcsec, and 60 arcsec, respectively.  Because the PA dispersion angular resolution was the largest of the three, the other two maps needed to be changed to this resolution.  The velocity dispersion and volume density maps were regridded to the resolution of the PA dispersion map, with 60$\times$60 arcsec pixels, where the value of each output pixel was the average of the values of the input pixels that overlapped the output pixel area, weighted by the fraction of the input pixel area falling into the output pixel.}

\begin{figure*}
				\begin{center} 
					\includegraphics[width = 6in]{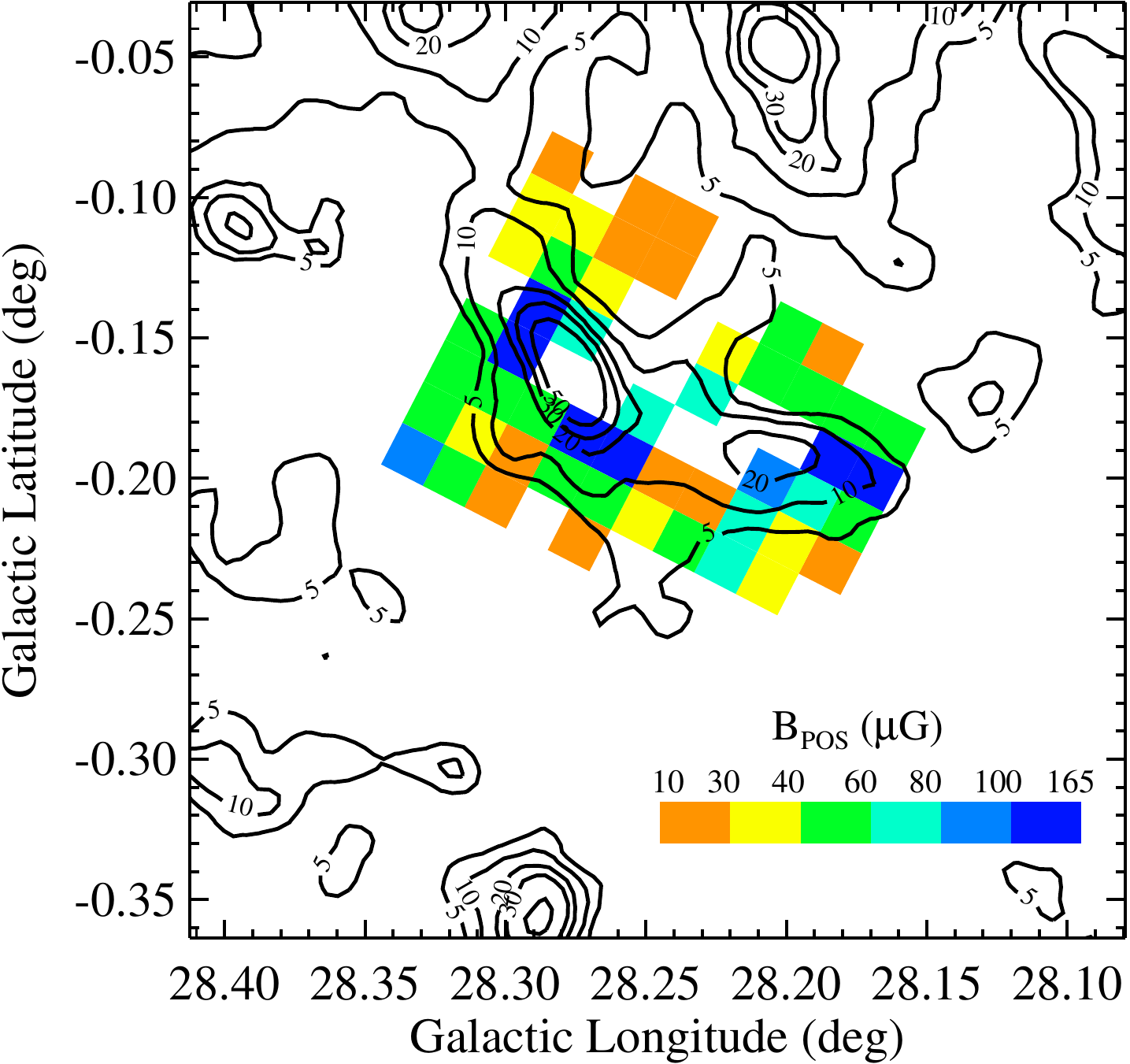}
					\caption{Plane-of-sky B-field strength for G28.23, estimated using the CF Method.  Black solid lines show the column-density-derived A$_V$ contours.  Filled pixels show the B$_{POS}$ values, as indicated by the color bar.  Missing pixels failed to meet the PA dispersion and/or volume density criteria.}
					\label{figbposstrength}
				\end{center}
			\end{figure*}

			\replaced{Equation \ref{eqnbpos} was used to calculate the B$_{POS}$ map of G28.23, shown in Figure \ref{figbposstrength}.}{Equation \ref{eqnbpos} was used to calculate B$_{POS}$ in each of these matched pixels and combined into a map of G28.23.  The matched maps consisted of 11$\times$11 pixels.  Of these, B$_{POS}$ was calculated in the 51 pixels where PA dispersions, minus their uncertainties, were less than or equal to 25$\degr$ and where the volume densities plus their uncertainties were greater than or equal to 50~H$_2$~cm$^{-3}$.  Uncertainties in the B$_{POS}$ estimates were found by propagating the uncertainties from the PA dispersions, velocity dispersions, and volume densities.  We assumed no uncertainty in the kinematic distance of 5.1 kpc.  The dependence of the B-field strength on distance (and any distance uncertainty) is discussed in Section \ref{systematicuncertaintiesall}.  The B$_{POS}$ SNR values ranged from 1.4 to 6.7, with a median of 4.2.}  
			
			The properties of each bin, including PA dispersion, volume density, $^{13}$CO gas velocity dispersion, B$_{POS}$, and the normalized Mass-to-Flux ratio (discussed in the following section), along with their uncertainties, are listed in Table \ref{table3}.  \deleted{The PA dispersion map consisted of 11$\times$11 pixels.  Of these, B$_{POS}$ was calculated in the 51 pixels where PA dispersions minus their uncertainties were less than or equal to 25$\degr$ and where the volume densities plus their uncertainties were greater than or equal to 50~H$_2$~cm$^{-3}$.}  The last column of Table \ref{table3} lists whether the PA dispersion, density, or both limited the B$_{POS}$ calculated for each pixel.  \replaced{A}{If the} PA dispersion of a pixel \deleted{was categorized as a limit if the dispersion} \replaced{is}{was} above 25$\degr$, but the dispersion minus its uncertainty \replaced{is}{was} less than or equal to 25$\degr$, \added{then the resulting B-field strength calculation yielded an upper limit}.  Similarly, \added{if} the volume density of a pixel \replaced{limited B$_{POS}$ if the density itself is}{was} below 50~H$_2$~cm$^{-3}$, but the density plus its uncertainty \replaced{is}{was} greater than or equal to 50~H$_2$~cm$^{-3}$, \added{then the resulting B-field strength was also an upper limit}.
			
                        \onecolumngrid
                        \newpage
			\begin{deluxetable}{cccccccc}
				\tabletypesize{\footnotesize}
				\tablecolumns{8}
				\tablecaption{Properties of the Pixels in the B-field Map}
				\tablewidth{0pt}
				\tablenum{3}
				\colnumbers
				\tablehead{
					\colhead{$l$} &
					\colhead{$b$} &
					\colhead{PA Dispersion} &
					\colhead{Volume Density} &
					\colhead{$\sigma_v$ ($^{13}$CO)} &
					\colhead{B$_{POS}$} &
					\colhead{\Mphip} &
					\colhead{Limits\tablenotemark{a}} \\
					\colhead{($^{\circ}$)} &
					\colhead{($^{\circ}$)} &
					\colhead{(deg)}        &
					\colhead{(H$_2$ cm$^{-3}$)} &
					\colhead{(km s$^{-1}$)}     &
					\colhead{($\mu$G)}     &
					\colhead{(Normalized)} &
					\colhead{Used}
				}
				\startdata
				28.1315 & -0.1638 & $\cdots$ & $\cdots$ & 2.1 (0.5) & $\cdots$ & $\cdots$ & $\cdots$ \\ 
				28.1463 & -0.1562 & $\cdots$ & $\cdots$ & 2.2 (0.3) & $\cdots$ & $\cdots$ & $\cdots$ \\ 
				28.1612 & -0.1486 & 25.9 (3.9) & $\cdots$ & 2.7 (0.2) & $\cdots$ & $\cdots$ & $\cdots$ \\ 
				28.1760 & -0.1410 & 19.7 (2.9) & $\cdots$ & 3.0 (0.2) & $\cdots$ & $\cdots$ & $\cdots$ \\ 
				28.1908 & -0.1334 & 11.8 (3.2) & $\cdots$ & 2.7 (0.5) & $\cdots$ & $\cdots$ & $\cdots$ \\ 
				28.2056 & -0.1258 & 26.1 (3.1) & $\cdots$ & 2.4 (0.2) & $\cdots$ & $\cdots$ & $\cdots$ \\ 
				28.2204 & -0.1181 & 26.1 (2.7) & 20 (9) & 2.9 (0.4) & $\cdots$ & $\cdots$ & $\cdots$ \\ 
				28.2353 & -0.1105 & 18.9 (3.0) & 59 (25) & 2.0 (1.8) & 17.5 (4.9) & 1.10 (0.33) & $\cdots$ \\ 
				28.2501 & -0.1029 & 13.6 (1.9) & 57 (24) & 2.4 (0.1) & 29.6 (7.4) & 0.78 (0.21) & $\cdots$ \\ 
				28.2649 & -0.0953 & 16.0 (2.3) & 15 (8) & 2.3 (0.1) & $\cdots$ & $\cdots$ & $\cdots$ \\ 
				28.2797 & -0.0877 & 24.8 (5.0) & 55 (25) & 2.4 (0.1) & 15.6 (4.8) & 2.88 (0.93) & $\cdots$ \\ 
				28.1391 & -0.1787 & $\cdots$ & 12 (5) & 3.1 (0.2) & $\cdots$ & $\cdots$ & $\cdots$ \\ 
				28.1539 & -0.1711 & $\cdots$ & 27 (15) & 2.5 (0.2) & $\cdots$ & $\cdots$ & $\cdots$ \\ 
				28.1688 & -0.1634 & 20.1 (2.9) & 31 (17) & 1.7 (0.2) & $\cdots$ & $\cdots$ & $\cdots$ \\ 
				28.1836 & -0.1558 & 14.1 (2.1) & 40 (19) & 1.4 (0.2) & 13.5 (3.8) & 1.51 (0.46) & n \\ 
				28.1984 & -0.1482 & 9.8 (2.3) & 41 (21) & 3.8 (5.4) & 54.8 (20.6) & 0.49 (0.19) & n \\ 
				28.2132 & -0.1406 & 25.8 (2.8) & 8 (5) & 8.4 (0.4) & $\cdots$ & $\cdots$ & $\cdots$ \\ 
				28.2281 & -0.1330 & 26.3 (2.6) & 28 (10) & 13.3 (1.3) & $\cdots$ & $\cdots$ & $\cdots$ \\ 
				28.2429 & -0.1254 & 16.5 (3.0) & 92 (31) & 1.9 (2.0) & 24.9 (6.8) & 0.62 (0.18) & $\cdots$ \\ 
				28.2577 & -0.1177 & 16.2 (2.3) & 100 (33) & 1.6 (0.1) & 22.0 (4.8) & 0.57 (0.14) & $\cdots$ \\ 
				28.2725 & -0.1101 & 14.6 (2.1) & 90 (35) & 2.7 (0.1) & 39.5 (9.5) & 0.78 (0.20) & $\cdots$ \\ 
				28.2874 & -0.1025 & 19.6 (3.4) & 111 (37) & 2.5 (0.1) & 30.1 (7.2) & 2.51 (0.65) & $\cdots$ \\ 
				28.1467 & -0.1935 & $\cdots$ & 117 (42) & 3.6 (0.3) & $\cdots$ & $\cdots$ & $\cdots$ \\ 
				28.1616 & -0.1859 & 20.7 (3.3) & 132 (44) & 3.4 (0.2) & 42.0 (9.7) & 1.26 (0.32) & $\cdots$ \\ 
				28.1764 & -0.1783 & 18.9 (2.7) & 122 (40) & 3.3 (0.3) & 42.6 (9.3) & 1.36 (0.33) & $\cdots$ \\ 
				28.1912 & -0.1706 & 27.3 (2.9) & 117 (39) & 5.9 (0.5) & 51.5 (10.3) & 0.50 (0.11) & PA \\ 
				28.2060 & -0.1630 & 26.5 (2.8) & 100 (34) & 5.3 (0.7) & 43.7 (8.8) & 0.28 (0.06) & PA \\ 
				28.2209 & -0.1554 & 16.6 (2.3) & 68 (28) & 3.4 (0.5) & 36.9 (9.1) & 1.56 (0.41) & $\cdots$ \\ 
				28.2357 & -0.1478 & 30.6 (1.8) & 39 (17) & 2.9 (4.2) & $\cdots$ & $\cdots$ & $\cdots$ \\ 
				28.2505 & -0.1402 & 29.0 (2.1) & 142 (43) & 1.3 (0.1) & $\cdots$ & $\cdots$ & $\cdots$ \\ 
				28.2653 & -0.1326 & 20.0 (2.8) & 190 (54) & 2.2 (0.1) & 33.7 (6.8) & 1.25 (0.28) & $\cdots$ \\ 
				28.2802 & -0.1249 & 18.1 (2.2) & 217 (63) & 2.3 (0.1) & 40.9 (7.8) & 1.84 (0.39) & $\cdots$ \\ 
				28.2950 & -0.1173 & 18.3 (2.0) & 249 (71) & 2.1 (0.1) & 39.4 (10.7) & 3.69 (1.06) & $\cdots$ \\ 
				28.1544 & -0.2083 & $\cdots$ & 404 (87) & 2.9 (0.6) & $\cdots$ & $\cdots$ & $\cdots$ \\ 
				28.1692 & -0.2007 & 14.7 (2.6) & 461 (96) & 3.2 (0.2) & 102.1 (21.1) & 1.18 (0.27) & $\cdots$ \\ 
				28.1840 & -0.1931 & 15.0 (2.0) & 521 (108) & 3.5 (0.3) & 116.6 (20.0) & 1.71 (0.34) & $\cdots$ \\ 
				28.1988 & -0.1855 & 29.0 (3.1) & 469 (106) & 6.4 (0.5) & $\cdots$ & $\cdots$ & $\cdots$ \\ 
				28.2137 & -0.1779 & 31.5 (3.6) & 283 (82) & 5.7 (0.6) & $\cdots$ & $\cdots$ & $\cdots$ \\ 
				28.2285 & -0.1702 & 17.7 (2.2) & 147 (45) & 4.5 (0.5) & 68.1 (13.6) & 1.76 (0.39) & $\cdots$ \\ 
				28.2433 & -0.1626 & 32.2 (1.9) & 121 (38) & 4.1 (0.5) & $\cdots$ & $\cdots$ & $\cdots$ \\ 
				28.2581 & -0.1550 & 37.2 (2.4) & 261 (72) & 1.9 (0.1) & $\cdots$ & $\cdots$ & $\cdots$ \\ 
				28.2730 & -0.1474 & 16.4 (4.1) & 516 (134) & 2.5 (0.1) & 76.0 (15.6) & 4.91 (1.12) & $\cdots$ \\ 
				28.2878 & -0.1398 & 8.6 (2.6) & 888 (182) & 2.1 (0.1) & 163.1 (24.4) & 3.17 (0.57) & $\cdots$ \\ 
				28.3026 & -0.1322 & $\cdots$ & 672 (135) & 1.9 (0.1) & $\cdots$ & $\cdots$ & $\cdots$ \\ 
				28.1620 & -0.2231 & $\cdots$ & 191 (57) & 2.3 (1.7) & $\cdots$ & $\cdots$ & $\cdots$ \\ 
				28.1768 & -0.2155 & 14.4 (3.0) & 155 (50) & 2.3 (0.2) & 44.1 (11.8) & 1.79 (0.51) & $\cdots$ \\ 
				28.1916 & -0.2079 & 15.8 (2.0) & 287 (79) & 3.3 (0.2) & 77.7 (14.6) & 1.63 (0.35) & $\cdots$ \\ 
				28.2064 & -0.2003 & 14.7 (2.7) & 473 (108) & 3.1 (0.2) & 99.9 (21.5) & 2.36 (0.56) & $\cdots$ \\ 
				28.2213 & -0.1927 & $\cdots$ & 610 (125) & 2.4 (0.5) & $\cdots$ & $\cdots$ & $\cdots$ \\ 
				28.2361 & -0.1851 & 29.5 (3.0) & 535 (110) & 2.6 (0.8) & $\cdots$ & $\cdots$ & $\cdots$ \\ 
				28.2509 & -0.1774 & 24.9 (2.7) & 508 (113) & 3.5 (1.2) & 68.8 (11.0) & 2.97 (0.56) & $\cdots$ \\ 
				28.2657 & -0.1698 & $\cdots$ & 835 (174) & 3.0 (0.0) & $\cdots$ & $\cdots$ & $\cdots$ \\ 
				28.2806 & -0.1622 & $\cdots$ & 1192 (238) & 2.6 (0.1) & $\cdots$ & $\cdots$ & $\cdots$ \\ 
				28.2954 & -0.1546 & 12.2 (3.0) & 649 (158) & 2.7 (0.1) & 123.8 (25.8) & 3.54 (0.82) & $\cdots$ \\ 
				28.3102 & -0.1470 & 16.5 (2.9) & 271 (78) & 2.3 (0.1) & 49.8 (11.4) & 2.48 (0.62) & $\cdots$ \\ 
				28.1696 & -0.2380 & $\cdots$ & 43 (19) & 2.0 (0.2) & $\cdots$ & $\cdots$ & n \\ 
				28.1844 & -0.2304 & 21.2 (3.3) & 46 (23) & 4.2 (0.3) & 29.6 (8.6) & 0.90 (0.28) & n \\ 
				28.1992 & -0.2227 & 19.0 (2.5) & 96 (34) & 3.2 (0.2) & 36.0 (8.0) & 1.35 (0.33) & $\cdots$ \\ 
				28.2141 & -0.2151 & 12.9 (1.7) & 127 (40) & 3.1 (0.2) & 60.3 (12.4) & 1.55 (0.35) & $\cdots$ \\ 
				28.2289 & -0.2075 & 27.1 (3.3) & 161 (50) & 1.6 (0.2) & 16.6 (3.3) & 6.40 (1.42) & PA \\ 
				28.2437 & -0.1999 & 26.3 (3.3) & 271 (75) & 1.2 (1.9) & 16.6 (4.0) & 7.51 (1.97) & PA \\ 
				28.2585 & -0.1923 & 11.8 (2.0) & 485 (112) & 2.9 (8.5) & 120.2 (42.6) & 1.39 (0.51) & $\cdots$ \\ 
				28.2734 & -0.1847 & 12.5 (2.8) & 601 (141) & 2.9 (0.1) & 125.8 (31.9) & 2.81 (0.77) & $\cdots$ \\ 
				28.2882 & -0.1770 & 26.3 (2.2) & 399 (107) & 3.0 (0.1) & 49.8 (7.9) & 6.32 (1.18) & PA \\ 
				28.3030 & -0.1694 & 26.4 (2.3) & 242 (67) & 3.7 (0.1) & 48.5 (7.9) & 2.68 (0.51) & PA \\ 
				28.3178 & -0.1618 & 15.4 (3.5) & 113 (36) & 3.0 (0.2) & 45.0 (12.6) & 0.95 (0.28) & $\cdots$ \\ 
				28.1772 & -0.2528 & $\cdots$ & $\cdots$ & 3.0 (0.3) & $\cdots$ & $\cdots$ & $\cdots$ \\ 
				28.1920 & -0.2452 & 23.6 (3.4) & 4 (2) & 4.6 (0.5) & $\cdots$ & $\cdots$ & $\cdots$ \\ 
				28.2069 & -0.2376 & 19.8 (2.2) & 36 (19) & 4.5 (0.5) & 30.2 (9.0) & 0.09 (0.03) & n \\ 
				28.2217 & -0.2299 & 12.3 (1.3) & 52 (23) & 5.7 (0.8) & 73.3 (18.0) & 0.45 (0.12) & $\cdots$ \\ 
				28.2365 & -0.2223 & 10.1 (1.5) & 61 (27) & 3.0 (0.3) & 50.7 (13.7) & 1.23 (0.36) & $\cdots$ \\ 
				28.2513 & -0.2147 & 12.5 (2.5) & 90 (33) & 2.2 (0.2) & 36.3 (9.9) & 2.17 (0.63) & $\cdots$ \\ 
				28.2662 & -0.2071 & 13.7 (2.2) & 122 (41) & 2.9 (1.0) & 51.2 (12.0) & 2.09 (0.53) & $\cdots$ \\ 
				28.2810 & -0.1995 & 23.4 (2.8) & 171 (53) & 3.5 (0.2) & 43.5 (8.5) & 2.73 (0.60) & $\cdots$ \\ 
				28.2958 & -0.1919 & 33.6 (9.5) & 155 (46) & 3.1 (0.1) & 25.2 (8.0) & 4.11 (1.37) & PA \\ 
				28.3106 & -0.1842 & 28.1 (8.9) & 135 (40) & 3.5 (0.2) & 32.1 (11.2) & 3.06 (1.11) & PA \\ 
				28.3255 & -0.1766 & 25.6 (3.2) & 80 (30) & 6.2 (0.5) & 47.7 (10.8) & 0.50 (0.12) & PA \\ 
				28.1848 & -0.2676 & $\cdots$ & $\cdots$ & 6.2 (0.7) & $\cdots$ & $\cdots$ & $\cdots$ \\ 
				28.1996 & -0.2600 & 16.7 (4.2) & $\cdots$ & 8.3 (0.8) & $\cdots$ & $\cdots$ & $\cdots$ \\ 
				28.2145 & -0.2524 & 17.5 (2.3) & $\cdots$ & 5.3 (0.5) & $\cdots$ & $\cdots$ & $\cdots$ \\ 
				28.2293 & -0.2448 & 13.5 (1.8) & $\cdots$ & 5.4 (0.5) & $\cdots$ & $\cdots$ & $\cdots$ \\ 
				28.2441 & -0.2372 & 10.6 (2.4) & 1 (0) & 4.2 (0.5) & $\cdots$ & $\cdots$ & $\cdots$ \\ 
				28.2589 & -0.2295 & 15.3 (2.7) & 22 (11) & 3.4 (0.6) & $\cdots$ & $\cdots$ & $\cdots$ \\ 
				28.2738 & -0.2219 & 18.2 (2.7) & 54 (24) & 3.0 (0.2) & 26.6 (7.1) & 1.45 (0.41) & $\cdots$ \\ 
				28.2886 & -0.2143 & 34.5 (4.4) & 90 (31) & 3.3 (0.1) & $\cdots$ & $\cdots$ & $\cdots$ \\ 
				28.3034 & -0.2067 & 34.9 (17.9) & 87 (30) & 2.7 (0.1) & 15.6 (8.5) & 1.19 (0.65) & PA \\ 
				28.3182 & -0.1991 & 22.1 (15.5) & 84 (29) & 5.2 (0.3) & 47.8 (34.5) & 0.61 (0.45) & $\cdots$ \\ 
				28.3331 & -0.1915 & 24.4 (4.9) & 67 (25) & 11.7 (0.7) & 86.8 (23.9) & 0.31 (0.09) & $\cdots$ \\ 
				28.1924 & -0.2824 & $\cdots$ & $\cdots$ & 9.0 (0.6) & $\cdots$ & $\cdots$ & $\cdots$ \\ 
				28.2073 & -0.2748 & 27.0 (3.8) & $\cdots$ & 11.2 (1.1) & $\cdots$ & $\cdots$ & $\cdots$ \\ 
				28.2221 & -0.2672 & 25.7 (2.6) & $\cdots$ & 6.6 (0.4) & $\cdots$ & $\cdots$ & $\cdots$ \\ 
				28.2369 & -0.2596 & 25.3 (2.6) & $\cdots$ & 6.1 (0.5) & $\cdots$ & $\cdots$ & $\cdots$ \\ 
				28.2517 & -0.2520 & 11.1 (2.6) & $\cdots$ & 3.2 (0.2) & $\cdots$ & $\cdots$ & $\cdots$ \\ 
				28.2666 & -0.2444 & 21.5 (2.6) & $\cdots$ & 2.8 (0.2) & $\cdots$ & $\cdots$ & $\cdots$ \\ 
				28.2814 & -0.2367 & 26.2 (3.0) & 4 (2) & 2.2 (0.1) & $\cdots$ & $\cdots$ & $\cdots$ \\ 
				28.2962 & -0.2291 & $\cdots$ & 49 (24) & 2.2 (0.1) & $\cdots$ & $\cdots$ & n \\ 
				28.3110 & -0.2215 & $\cdots$ & 55 (29) & 1.8 (0.1) & $\cdots$ & $\cdots$ & $\cdots$ \\ 
				28.3259 & -0.2139 & $\cdots$ & 41 (20) & 8.0 (1.5) & $\cdots$ & $\cdots$ & n \\ 
				28.3407 & -0.2063 & $\cdots$ & 47 (23) & 7.9 (12.4) & $\cdots$ & $\cdots$ & n \\ 
				28.2000 & -0.2973 & $\cdots$ & $\cdots$ & 3.9 (0.3) & $\cdots$ & $\cdots$ & $\cdots$ \\ 
				28.2149 & -0.2896 & 33.2 (3.2) & $\cdots$ & 5.3 (0.7) & $\cdots$ & $\cdots$ & $\cdots$ \\ 
				28.2297 & -0.2820 & 32.5 (2.7) & 0 (0) & 4.9 (0.4) & $\cdots$ & $\cdots$ & $\cdots$ \\ 
				28.2445 & -0.2744 & 31.8 (4.2) & 0 (0) & 10.4 (0.8) & $\cdots$ & $\cdots$ & $\cdots$ \\ 
				28.2593 & -0.2668 & $\cdots$ & 0 (0) & 6.8 (0.6) & $\cdots$ & $\cdots$ & $\cdots$ \\ 
				28.2742 & -0.2592 & 22.8 (3.8) & $\cdots$ & 2.3 (0.1) & $\cdots$ & $\cdots$ & $\cdots$ \\ 
				28.2890 & -0.2516 & 20.6 (3.6) & $\cdots$ & 2.0 (0.1) & $\cdots$ & $\cdots$ & $\cdots$ \\ 
				28.3038 & -0.2440 & $\cdots$ & $\cdots$ & 1.6 (0.1) & $\cdots$ & $\cdots$ & $\cdots$ \\ 
				28.3187 & -0.2363 & $\cdots$ & $\cdots$ & 1.7 (0.1) & $\cdots$ & $\cdots$ & $\cdots$ \\ 
				28.3335 & -0.2287 & $\cdots$ & $\cdots$ & 7.9 (18.3) & $\cdots$ & $\cdots$ & $\cdots$ \\ 
				28.3483 & -0.2211 & $\cdots$ & $\cdots$ & 6.5 (14.3) & $\cdots$ & $\cdots$ & $\cdots$ \\ 
				28.2077 & -0.3121 & $\cdots$ & $\cdots$ & 8.7 (0.8) & $\cdots$ & $\cdots$ & $\cdots$ \\ 
				28.2225 & -0.3045 & $\cdots$ & $\cdots$ & 5.3 (0.5) & $\cdots$ & $\cdots$ & $\cdots$ \\ 
				28.2373 & -0.2968 & $\cdots$ & $\cdots$ & 6.9 (0.8) & $\cdots$ & $\cdots$ & $\cdots$ \\ 
				28.2521 & -0.2892 & $\cdots$ & $\cdots$ & 14.9 (1.8) & $\cdots$ & $\cdots$ & $\cdots$ \\ 
				28.2670 & -0.2816 & $\cdots$ & $\cdots$ & 11.4 (1.1) & $\cdots$ & $\cdots$ & $\cdots$ \\ 
				28.2818 & -0.2740 & $\cdots$ & $\cdots$ & 4.3 (0.3) & $\cdots$ & $\cdots$ & $\cdots$ \\ 
				28.2966 & -0.2664 & $\cdots$ & $\cdots$ & 2.6 (0.2) & $\cdots$ & $\cdots$ & $\cdots$ \\ 
				28.3114 & -0.2588 & $\cdots$ & $\cdots$ & 2.4 (0.2) & $\cdots$ & $\cdots$ & $\cdots$ \\ 
				28.3263 & -0.2512 & $\cdots$ & $\cdots$ & 2.9 (0.2) & $\cdots$ & $\cdots$ & $\cdots$ \\ 
				28.3411 & -0.2435 & $\cdots$ & $\cdots$ & 4.2 (0.3) & $\cdots$ & $\cdots$ & $\cdots$ \\ 
				28.3559 & -0.2359 & $\cdots$ & $\cdots$ & 1.8 (2.1) & $\cdots$ & $\cdots$ & $\cdots$ \\ 
				\enddata
				\tablenotetext{a}{Indicates whether the PA dispersion (PA) or density (n) used to calculate B$_{POS}$ limited the B$_{POS}$ calculation.  ``PA" indicates the PA dispersion was greater than 25$\degr$, but PA dispersion minus its uncertainty was less than or equal to 25$\degr$.  ``n" indicates that the volume density was less than 50~H$_2$~cm$^{-3}$, but the density plus its uncertainty was greater than or equal to 50~H$_2$~cm$^{-3}$.}
				\label{table3}
			\end{deluxetable}
						
\twocolumngrid						
					
			\deleted{As shown in} Figure \ref{figbposstrength} \added{presents the B-field strengths in map form, where} the derived field strengths range from 10--165~$\mu$G, and probe extinctions up to A$_V$$\sim$30 mag.  The field strength estimates are predominantly \replaced{lowest in the}{lower in the lower density regions of the} upper left (NE) \replaced{region}{quadrant} of Figure \ref{figbposstrength}.  The field strength could not be probed in the densest regions of the cloud because of the lack of \added{NIR background} stars with polarimetric \replaced{measurements}{detections}.  
			
			\deleted{Uncertainties in the B$_{POS}$ estimates were found by propagating the uncertainties from the PA dispersions, velocity dispersions, and volume densities.  The B$_{POS}$ SNR ranged from 1.4 to 6.7, with a median of 4.2.}

			\subsubsection{B-field Strength versus Density}
\begin{figure*}  
				\begin{center}
					\includegraphics[width = 6in]{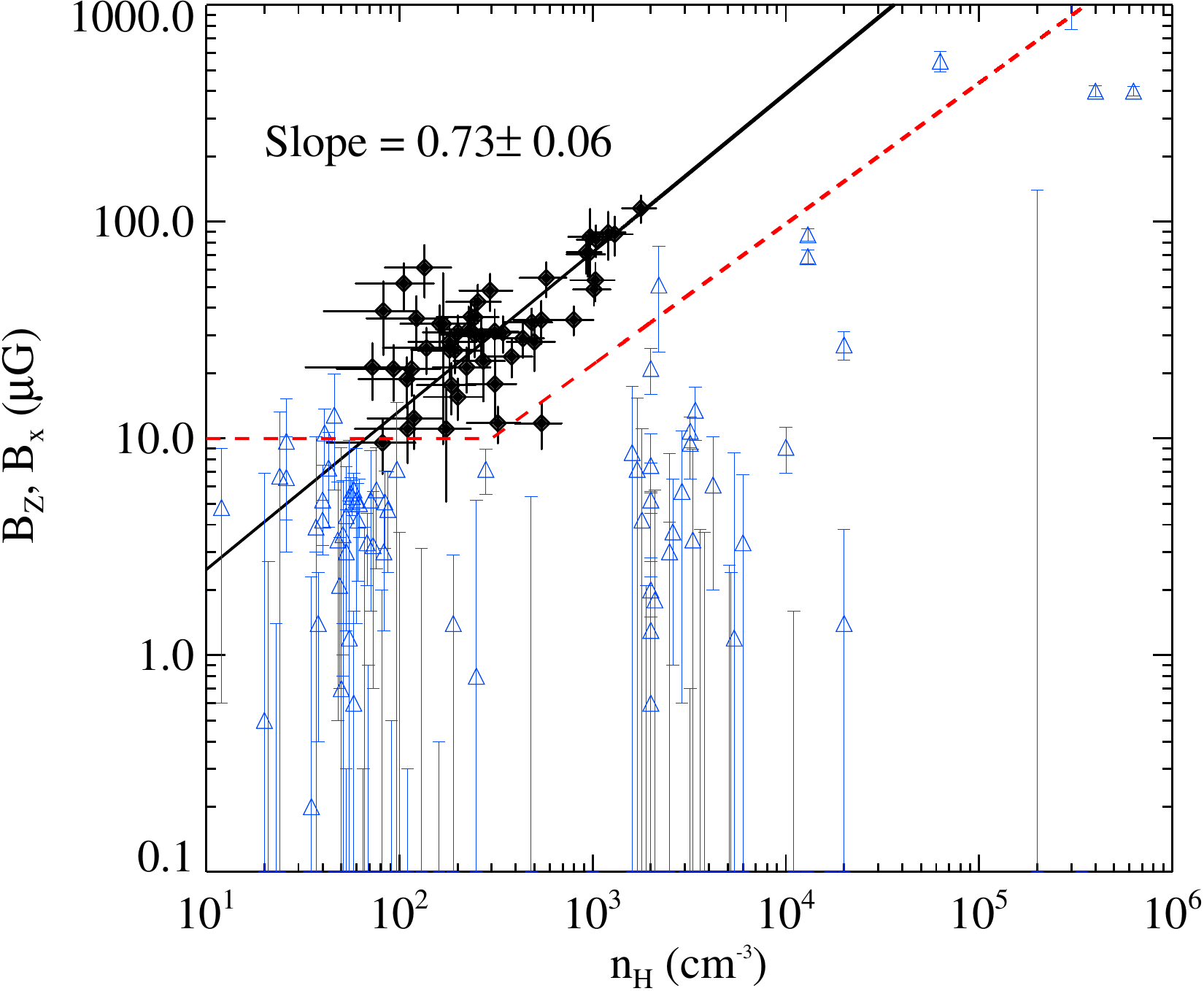}
					\caption{The plane-of-sky B-field strengths of the 51 pixels across G28.23, scaled by 1/$\sqrt{2}$ to become B$_x$, plotted against atomic hydrogen volume density, as the black diamonds.  B$_z$ estimates and upper limits from \citet{Crutcher2010} are shown as the blue triangles with corresponding uncertainties.  The \citet{Crutcher2010} maximum B-field estimate, B$_{max}$, is overlaid as the red, broken, dashed line.  A single power law fit to the B$_{x}$ values of G28.23 is shown as the black solid line, which has been extrapolated to densities outside those probed toward G28.23.}
					\label{figbvsvolden}
				\end{center}
			\end{figure*}

			One test of whether \replaced{the}{a} B-field \added{could have} influenced the flow of material into a cloud is \added{determining the} dependence of the field strength on cloud density.  This relation between B-field strength and average volume density is shown for G28.23 in Figure \ref{figbvsvolden}, which is based on, and uses data from, Figure 1 of \citet{Crutcher2010}.  The B$_{POS}$ strengths of G28.23 were scaled to represent the mean B-field strength in only one dimension ($B_x = B_{POS}/\sqrt{2}$), and are shown as the black diamonds.  The volume density ($n$) used in this relation is the volume density of atomic hydrogen, found by doubling the molecular hydrogen volume density.  The Zeeman measurements and upper limits from \citet{Crutcher2010} are shown as the blue triangles.  \replaced{along with}{Also shown is a red, dashed line representing} the equation of B$_{max}$ from their Equation 21.  Most of the B$_x$ \replaced{points}{values} for G28.23 \replaced{are larger than}{exceed} the corresponding B$_{max}$ values at the given densities.  We note that the B$_x$ points of G28.23 are for a single cloud, while the B$_Z$ points of \citet{Crutcher2010} consist of one point per cloud for various samples of clouds and Zeeman probes.
		
			A single power law of the form $B{\sim}n^{\alpha}$ was fit to the G28.23 B$_x$ versus $n$ points.  The best-fit power law had a slope $\alpha$=0.73~$\pm$~0.06 (black solid line in Figure \ref{figbvsvolden}).  \replaced{, which has a linear correlation coefficient of 0.64.}{The implications of this result will be discussed in Section \ref{bfielddensitymfratio}.}
		
		
		\subsection{Mass-to-Flux Ratio}
			The \Mphi~of a region determines whether gravitational or magnetic energy dominates \citep{Crutcher2012}.  \deleted{We calculated the normalized \Mphi~of the regions of G28.23 where B$_{POS}$ values were calculated (filled pixels in Figure \ref{figbposstrength}).}  The normalized \Mphi~(\Mphin), is equal to \citep{Crutcher2004}:
		
			\begin{equation}
			M/\Phi_{BN} = 7.6 \times 10^{-21} N_{H_2}/ B,
			\end{equation}
			where N$_{H_2}$ is in cm$^{-2}$ and B is in $\mu$G.  A normalized \Mphi~equal to unity indicates that gravitational energy is equal to the magnetic energy.  
			
			\added{We calculated the normalized \Mphi~of the regions of G28.23 where B$_{POS}$ values were calculated (filled pixels in Figure \ref{figbposstrength}).}  The B-field strength used in the \Mphin~calculation is nominally B$_{TOT}$, which is the amplitude of the 3-D vector B-field strength \citep{Crutcher2004}.  However, because we measure B$_{POS}$, the \Mphin~calculated here is actually \Mphip~\citep{Planck2016b}, \added{with the input B-field strength being B$_{POS}$}.  Based on the geometry of the cloud and whether the B-field is perpendicular or parallel to the cloud major axis, the average \Mphin~over all possible inclination angles (with respect to the line of sight) will be \citep{Crutcher2004, Planck2016b}:
			
			\begin{equation}
			\overline{M/\Phi} = \int_{0}^{\pi/2} \frac{M_{\parallel}~\cos{\theta}}{\Phi_{\perp}/\sin{\theta}}\sin{\theta}d{\theta} = f(M_{\parallel}/\Phi_{\perp}),
			\end{equation}
			where $\theta$ is the inclination of the cloud with respect to the line of sight.  Limiting geometries include: the B-field being perpendicular to the cloud major axis, and, the B-field being parallel to the cloud major axis.  The correction factor, $f$, applied to \Mphip~to yield the average \Mphin, will be bounded by 1/3 (perpendicular) and 3/4 (parallel) \citep{Planck2016b}.  Therefore, the \Mphip~values presented for G28.23 need correction, depending on geometry, to \replaced{predict}{infer} \Mphin.  \added{This correction does assume that all inclination angles are likely, which may not be the case.}
			 
\begin{figure*}  
				\begin{center}
					\includegraphics[width = 6in]{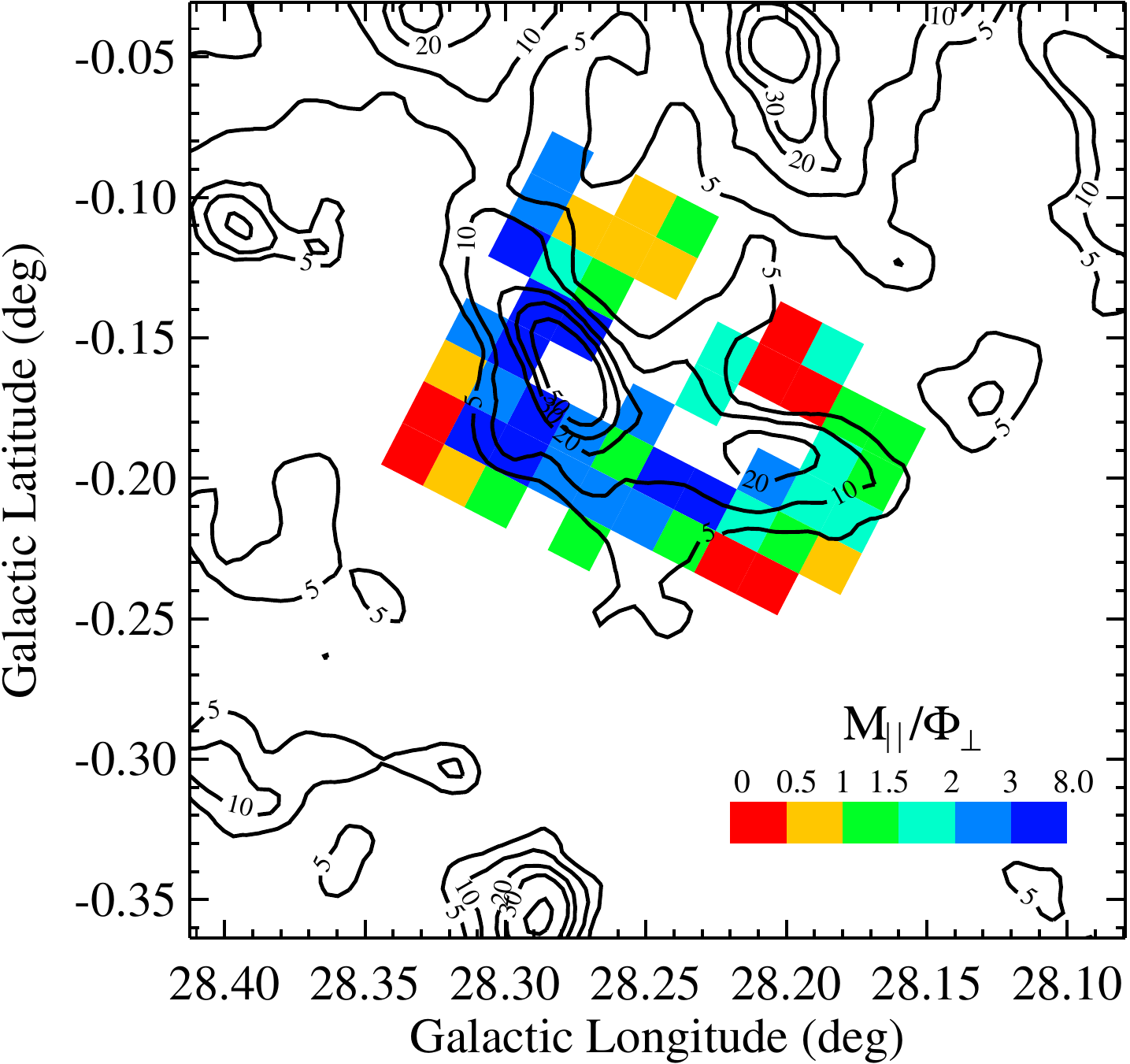}
					\caption{Normalized Mass-to-Flux ratios observed toward G28.23, with values indicated by colors corresponding to the color bar.  Column-density-derived A$_V$ contours are overlaid in black.  \deleted{The lowest \Mphip~values are at the cloud edge, while the highest values are closer to the cloud center.}}
					\label{figmassfluxnorm}
				\end{center}
			\end{figure*}

			Figure \ref{figmassfluxnorm} shows the \Mphip~map of G28.23.  The \Mphip~ratios range from $\sim$0.09--7.5, with a median of $\sim$1.55.  The \deleted{trend of} \Mphip~values appear to increase with increasing cloud density.  \added{This increase can be seen in the figure, where the lowest \Mphip~values are at the cloud edge, while the highest values are closer to the cloud center.}
		
\begin{figure}  
				\begin{center}
					\includegraphics[width =  \columnwidth]{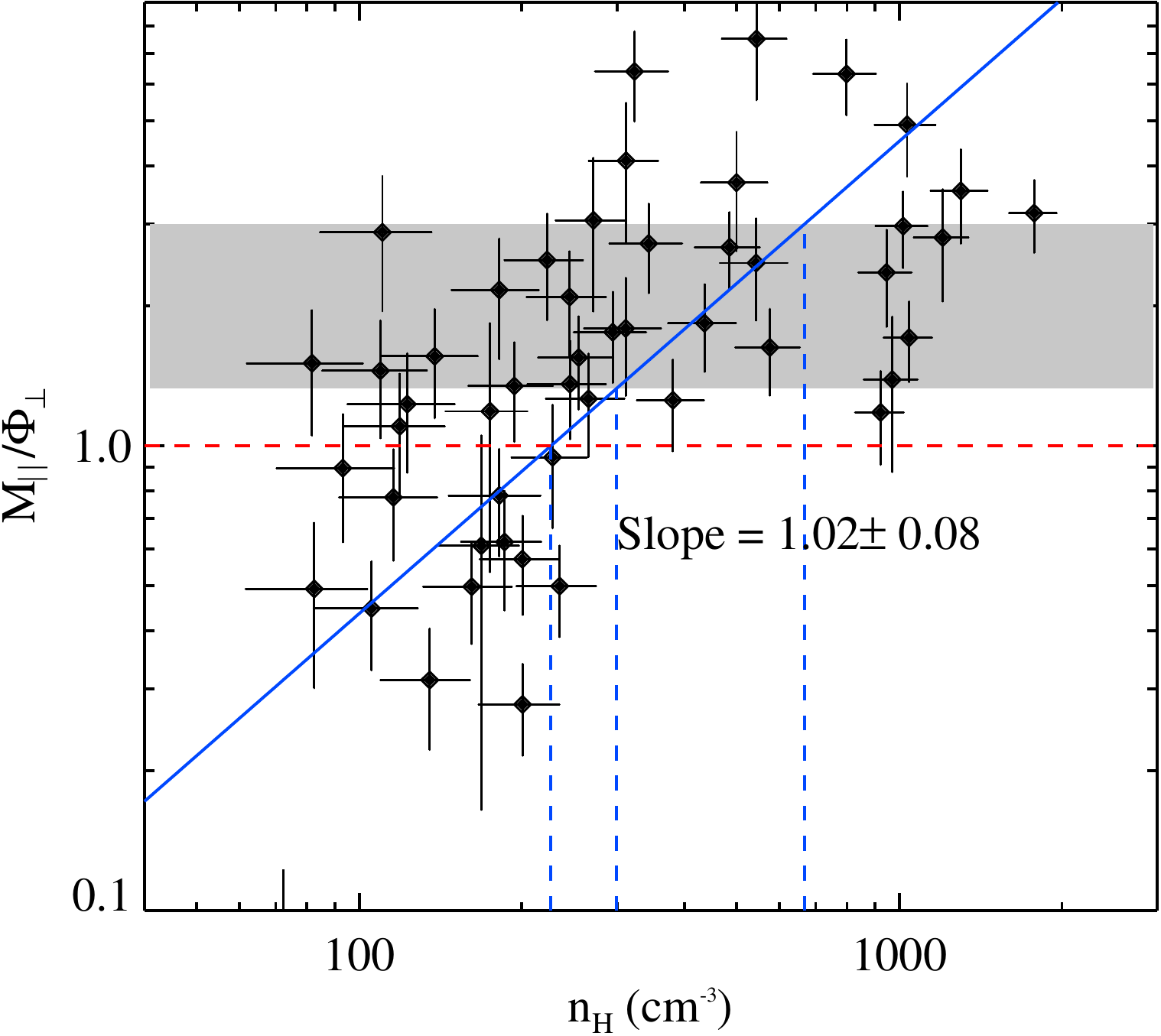}
					\caption{Normalized Mass-to-Flux ratio (\Mphip) values plotted against atomic hydrogen volume density, as black diamonds.  Similar to Figure \ref{figbvsvolden}, a single power law (blue) was fit to the data.  The horizontal dashed red line represents unity in uncorrected \Mphip.  The gray region represents unity in \Mphin~when \Mphip~is scaled by 1/3 to 3/4, depending on the geometry of the system.  The dashed vertical blue lines indicate the corresponding critical densities.}
					\label{figmfratiovsvolden}
				\end{center}
			\end{figure}

			Figure \ref{figmfratiovsvolden} plots the \Mphip~values against atomic hydrogen volume density.  The horizontal dashed line represents unity for \Mphip.  A positive correlation is seen between density and \Mphip, where regions that are less dense tend to have lower \Mphip~values, and regions that are denser tend to have larger \Mphip~values.  Applying the geometric correction factor lowers the \Mphip~estimated \Mphin~values and changes the average volume density at which the cloud goes from subcritical (B-field dominated) to supercritical (gravitationally dominated).  Similar to the B-field strength, a power law \added{of the form \Mphip~$\sim~n^{\alpha}$} was fit to \Mphip~versus $n$.  The best-fit power-law has an index of 1.02$\pm$0.08.  The uncorrected volume density that corresponds to criticality is 225~H~cm$^{-3}$, whereas the corrected \added{critical} volume density ranges from $\sim$300 to 670~H~cm$^{-3}$, depending on the correction factor.
		
	\section{Discussion}\label{Discussion}
		To test the importance of the B-field in IRDC G28.23, we examined the relationships between the B-field and other cloud properties.  In this section, we discuss the implications of the results.
		
		\subsection{Plane-of-sky B-field Morphology: Relative PA Orientations}
			The first test of the importance of the B-field in the formation of G28.23, Question 2, examines the relative orientation of the polarization PAs with respect to the projected cloud orientation.  For this IRDC, the relative PA orientations were neither preferentially perpendicular nor parallel to the cloud major axis orientation.  The relative PA orientations do, however, show a large-scale pattern, and appear to have distinct distributions in the Northern and Southern regions of the cloud, as seen in Figure \ref{figdelpasn25}, \added{with nearly a 30$\degr$ change in median relative PAs across the cloud spine location}.  The PAs in the North are more likely to be perpendicular to the cloud elongation, especially in the Northeast, whereas in the South \added{(especially Southwest)}, PAs are more likely to be parallel to the cloud elongation.  \deleted{Therefore, it is likely that the presence of the cloud has affected the B-field entrained in the region, but it is not clear to what extent the B-field influenced the cloud.}
			
			\added{Numerical simulations of filament formation in magnetized media \citep[e.g., ][]{Van2014} find that in regions where the B-field is strong, the orientation of the B-field is aligned with that of the cloud.  Recent studies of a nearby IRDC, G14.2 \citep{Busquet2013, Santos2016} and the relative orientations between {\it Planck} observations and nearby molecular clouds \citep{Planck2016a} have found that the B-field is preferentially perpendicular to dense cloud filaments.  Using {\it Herschel} observations of a portion of the Taurus Molecular cloud (B211 \& L1495), \citet{Palmeirim2013} found that less dense filaments, which are preferentially parallel to the surrounding B-field, connect perpendicularly to the main filament.}
			
			\replaced{The B-field morphology of G28.23 shares some similarities to a hub-filament system. In such a system,}{One specific cloud configuration in a strong B-field regime is the hub-filament system, where} the B-field helps funnel material to a central dense hub along one or more filaments \citep{Myers2009, Chapman2011, Li2014, Pillai2015}.  Such a configuration would \replaced{have}{exhibit} B-fields perpendicular to the dense hub and parallel to the less dense connecting filaments.  \deleted{The relative PA orientations in Figure \ref{figdelpasn25} suggest that may be the case for G28.23.  The PAs in the Eastern, denser, portion of the cloud, are preferentially perpendicular, especially in the Northern quadrant.  The Western portion, a less dense filament, shows a more parallel relative PA distribution.} \added{ In simulations of turbulent molecular clouds, \citet{Soler2013} also found that in cloud with high B-field strengths, the relative field orientations changed from parallel to perpendicular as a function of density.}  
			
\begin{figure*} 
				\includegraphics[width=6in]{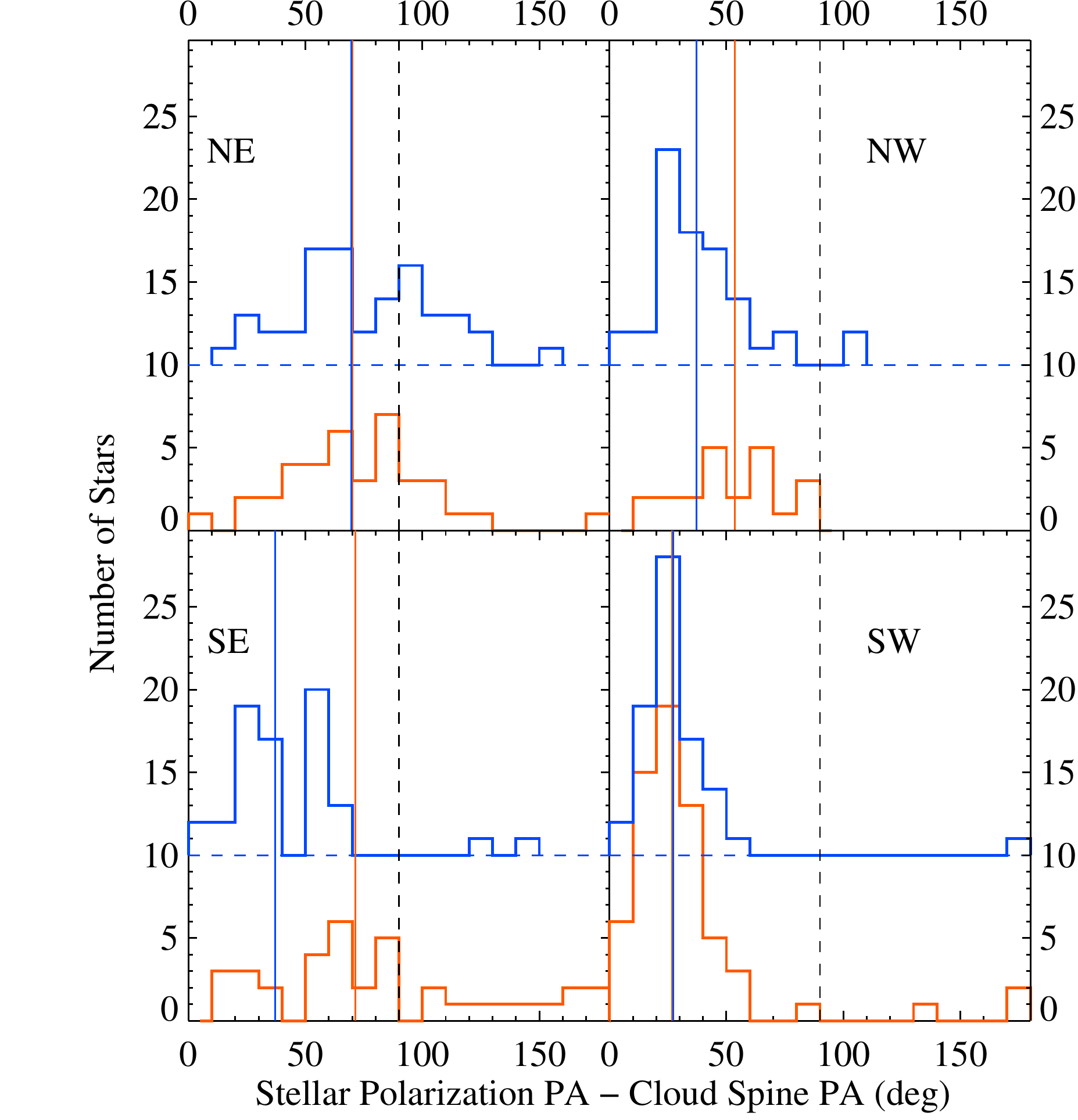}
				\caption{Similar to Figure \ref{figdelpasn25} but relative PA distributions separated into `near' (blue, offset by 10 counts for clarity) and `far' (orange) categories based on the average distance between the polarization stars and the cloud spine.  The locations of the medians of the distributions are indicated by the vertical color-coded lines.  In the Northwest and Southeast quadrants, the PAs closer to the cloud spine are more likely to be parallel to the cloud elongation.}
				\label{figdelpasn25nearfar}	
			\end{figure*}

\begin{figure*} 
				\includegraphics[width=6in]{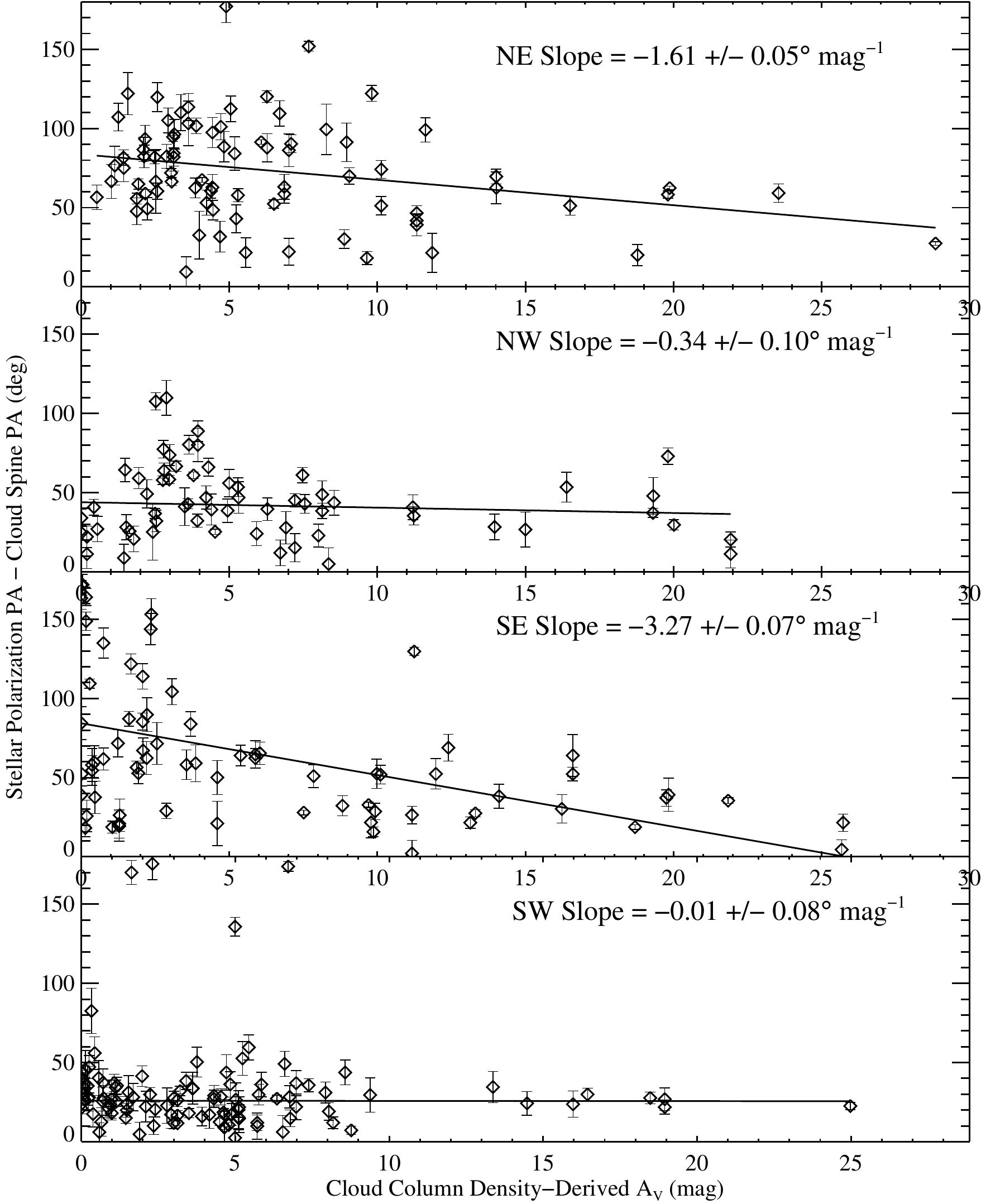}
				\caption{Relative stellar polarization PA versus the cloud column density-derived A$_V$ along the line of sight to each star.  The relations have been separated into the two quadrants designated in Figure \ref{figpaslicelinespine}.  The relative PAs tend to either remain parallel to the cloud axis \added{(NW, SW)} or change from preferentially perpendicular to parallel as a function of cloud column density \added{(NE, SE)}.}
				\label{figg2823delpavscold}
			\end{figure*}

			\added{To test this scenario, the relative PA orientations shown in Figure \ref{figdelpasn25} were separated into `close' and `far' subsamples by the median distance between the polarization stars and the cloud spine.  This median distance was $\sim$3.3 pc, and the median distances in the four quadrants ranged from $\sim$2.5 to 4.1 pc.  Figure \ref{figdelpasn25nearfar} shows that the relative PAs for the subsample of stars that are located closer to the cloud spine, and thus probe the higher density cloud regions, are more likely to be parallel to the cloud than the PAs for the subsample of stars that are farther from the cloud, and probe the less dense cloud regions.  The median angles of the near and far distributions in each quadrant are listed in the first portion of Table \ref{tablequadrantdist}, as well as the KS probability determining whether the two distributions were drawn from the same parent population.  
				
			The relations between the relative PAs and the cloud column density were also found.  Figure \ref{figg2823delpavscold} shows the relative PAs versus the cloud dust emission cloud column density-derived A$_V$ along the line of sight to each star.  The PAs shown here are the same ones shown in Figures \ref{figdelpasn25} and \ref{figdelpasn25nearfar}.  The relations were separated into the four cloud quadrants, and the PA variance-weighted best fitting line was found for each one.  Similar to the behavior seen in Figure \ref{figdelpasn25nearfar}, the relative PAs tend to become more parallel to the cloud elongation as a function of increasing column density.  This result is in contrast to what would be expected in a hub-filament system, where the nearer PAs would be expected to be perpendicular to the cloud.
				
			This relative PA configuration may be a signature of a helical field \citep{Fiege2000}.  In such a field, the B-field in the outer regions of the cloud would be toroidally dominated and the B-field in the denser regions of the cloud would be poloidally dominated.  It is not clear at what density or radius from the cloud spine the transition in the relative orientation is expected.  In a study of high-mass filaments, \citet{Contreras2013} found that their density distributions were consistent with that expected of a cloud wrapped by a helical field.  
				
			It is evident from images of G28.23, such as Figure \ref{figglimpseindvecbackforesub}, that the cloud spine is not one straight filament, but has some curvature.  Therefore, some of the changes in relative PA between the four quadrants could be due to the change in orientation of the filament along the cloud spine.  To test for this effect, we separated the Galactic PAs into the four quadrants and near and far distributions, very similar to the relative PAs.  If the cloud has affected the orientation of the local B-field, then the GPA distributions should differ between near and far subsamples.  The second portion of Table \ref{tablequadrantdist} lists the medians and KS probabilities between the near and far GPA distributions for each quadrant.  While the near and far distributions of the NE and SE quadrants have similar characteristics, the near and far distributions of the NW and SW quadrants are different.  For both quadrants, the near subsamples are more likely to be closer to 90$\degr$, or more plane-parallel.  }
			
                        \clearpage
			\begin{deluxetable}{cccc}
				\tabletypesize{\footnotesize}
				\tablecolumns{4}
				\tablecaption{PA Distribution Properties in the four Quadrants}
				\tablewidth{0pt}
				\tablenum{4}
				\colnumbers
				\tablehead{
					\colhead{Distribution} &
					\colhead{Near Median} &
					\colhead{Far Median} &
					\colhead{KS Prob.\tablenotemark{a}} \\
					\colhead{Quadrant} &
					\colhead{(deg)} &
					\colhead{(deg)} &
					\colhead{   }
			}
			\startdata
			$\Delta$PA NE &           69 &           70 &  0.83 \\
			$\Delta$PA NW &           37 &           53 &  0.02 \\
			$\Delta$PA SE &           37 &           71 &   2$\times$10$^{-6}$ \\
			$\Delta$PA SW &           27 &           26 &  0.95 \\
			\hline
			GPA NE &          104 &          110 &  0.43 \\
			GPA NW &          115 &          131 &  0.03 \\
			GPA SE &          107 &          104 &  0.39 \\
			GPA SW &           98 &          114 &  0.01 \\
			\enddata
			\tablenotetext{a}{Probability that the near and far distributions are drawn from the same parent population.}
			\label{tablequadrantdist}
			\end{deluxetable}

			Based on the relative PA orientations with respect to locations in the cloud, it is unclear whether the B-field played a dominant role in the formation of \replaced{the IRDC}{G28.23}.  
			
			We note the presence of a bright 8~$\mu$m feature in the Southeast region ($l\sim$28$\fdg$3, $b\sim-$0$\fdg$2) of the cloud, which may have disrupted the B-field in that area.  As seen in Figure \ref{figglimpseuqavgbackforesub}, the polarization measurements appear to form a shell around this bright ring-like structure.  Fewer high SNR stars are present near the structure, compared to the rest of the region.  This structure may correspond to a large bubble in the Catalog of bubbles from the Milky Way Project, centered at $l$ = 28$\fdg$297 and $b$ = $-$0$\fdg$202 \citep{Simpson2012}.  The bubble may have disrupted either the B-field or the dust grains in the region such that polarizations cannot easily be measured.  \replaced{As spectral}{Spectral} information targeting this structure was unavailable at the time of this study. \replaced{and it is not noticeable in the GRS $^{13}$CO data, the distance to this bubble is not known.}{While GRS $^{13}$CO line emission at different radial velocities toward this region (lines at 74 and 81 km s$^{-1}$) can be detected, this feature is not seen as a coherent structure in the $^{13}$CO data.  We conclude that the distance to this bubble is not known, and its association with G28.23 remains circumstantial.}
			
			\added{\subsubsection{Projection Effects}
			One concern in determining the relative angles between polarization measurements and cloud orientations are 3-dimensional projection effects.  Both the polarization and cloud orientations are projections onto the 2-dimensional plane of the sky of a 3-D B-field and cloud.  \citet{Hull2014}, using Monte Carlo simulations, explored projection effects in the context of protostellar outflow and B-field orientations.  They found that, unless the relative orientations are within 20 degrees of predominantly parallel, projection effects could produce seemingly perpendicular or other orientations.  
					
			Projection effects can thus affect the determination of whether the B-field dominated the formation of G28.23 based on the relative projected angles.  However, the relative angles between the polarization measurements themselves are unlikely to be affected by projection effects if they measure the same 3-dimensional B-field pointed in one direction.  Therefore, the shift in PAs between the Northern and Southern portions of the cloud, where the PAs are more likely to be perpendicular in the North and parallel in the South with respect to the cloud elongation, still indicates that the presence of the cloud changed the shape of the B-field.}
		
			\subsection{B-field, Cloud Density, and Mass-to-Flux}\label{bfielddensitymfratio}
			For the regions where B$_{POS}$ strengths were estimated for G28.23, the relation between B$_{x}$ and volume density was best fit with a single power law of slope 0.73$\pm$0.06, which addresses Question~3.  This slope is approximately equal to the slope of 0.65 found by \citet{Crutcher2010} for B vs. $n$ at densities larger than 300~H~cm$^{-3}$.
			
			Our best-fit slope, which is greater than 2/3, implies that the B-field was not the dominant force in the formation of G28.23.  However, the slope of 2/3 corresponds to a {\it temporal} evolution of the B-field with density as material undergoes collapse \citep{Li2015}.  The B vs. $n$ relation of G28.23 characterizes {\it instantaneous} structure: the present-day dependence of the B-field strength on density across the cloud.  \replaced{This information, if}{If} the temporal test of B-field importance can be applied to the structural relation, \added{then the B vs. $n$ relation} indicates that the B-field did not affect the formation of G28.23 to a large degree.  However, the influence of the B-field in the future evolution of the cloud is not known.  
			
			Simulations run by \citet{LKM2015} of clump collapse using both initially strong B-fields (Alfv\'{e}n Mach number $\mathcal{M}_{a}$$\sim$1) and weak B-fields ($\mathcal{M}_{a}$$\sim$10) found similar dependences of the B-field strength on density to those found in this study.  Their strong-field simulations exhibited a power law index of 0.7, which \added{was their time-averaged value of the index fit to the B vs. $n$ relation of a 100-cloud sample at times ranging from 0.4 to 0.64$t_{ff}$.  This value} is very similar to the \added{index of} 0.73 found here.  In contrast, their power law dependence in the weak-field simulation was 0.57.  The 2/3 dependence of \citet{Mestel1966} applied to weak B-fields, which is not the case for G28.23, but the similar dependence on density is still seen.  \added{The \citet{LKM2015} results imply that the 2/3 dependence on density can still arise in a molecular cloud with initially strong B-fields.}

			The median uncorrected \Mphip~of the cloud is $\sim$1.5, for which \Mphin~will range from 0.5 to 1.125 when corrected (Question 4).  \Mphip~was found to correlate with density with a power law index of $\sim$1.  This result indicates that there is a critical density, $n_{crit}$, above which the cloud is supercritical and no longer B-field supported.  The gray region in Figure \ref{figmfratiovsvolden} indicates where \Mphin~=~1 would be if a geometric correction factor is applied.  The best-fit line between density and \Mphip~intercepts this region for densities in the range $\sim$300-670~H~cm$^{-3}$.  The lower limit of this range is similar to the threshold density of 300~H~cm$^{-3}$ \replaced{at which}{that} \citet{Crutcher2010} \replaced{state that}{associated with} a molecular cloud \replaced{becomes}{becoming} self-gravitating.  \deleted{Our result implies that the onset of self-gravitation also corresponds to a departure of substantial B-field support of this cloud.}
			
			\added{The increase in \Mphip as a function of cloud density implies that the B-field is dynamically important, as this trend is a prediction of ambipolar diffusion \citep[e.g.,][]{Ciolek1994}.  This result contradicts the interpretation that the B-field did not play a dominant role in G28.23's formation based on the relative orientations of the polarization measurements and the power law dependence of the B-field strength on density.  
				
			Previous observations of the lifetimes of prestellar cores \citep[e.g.,][]{Ward-Thompson2007}, which are estimated to be on the order of a few times the free-fall time, are far shorter than those predicted by ambipolar diffusion models \citep[e.g.,][]{Mouschovias1999}.  To account for this discrepancy, it has been suggested that large-scale turbulence, which can create shocks that lead to regions of compressed gas, must be taken into account, in addition to ambipolar diffusion \citep[e.g.,][]{Li2004, Chen2014}.  In the case of G28.23, we speculate that the B-field may have initially been strong, and that ambipolar diffusion may have occurred in the cloud envelope as mass drifted into the cloud center, but the cloud is now gravitationally dominated in its denser regions.}
		
			\added{\subsection{Systematic Uncertainties of the B-field Strength}\label{systematicuncertaintiesall}
			The uncertainties reported in Table \ref{table3} for the B$_{POS}$ strengths are observational uncertainties, propagated from the random uncertainties of the volume density, the PA dispersion, and the $^{13}$CO gas velocity dispersion.  Systematic uncertainties will also affect the accuracy of the derived properties, and are likely larger than the observational uncertainties \citep[e.g.,][]{Crutcher2004}.  These systematics rise from assumptions made in calculating the B-field strength.  First, because the goal was to calculate the B-field strength in the plane of the sky, ideally, the gas velocity dispersion used in the CF calculation would be the velocity dispersion measured in the plane of the sky from tangential gas velocities.  However, as that information is not available, the dispersion along the line of sight was used as the only suitable proxy, though that imposes a velocity isotropy assumption that may not be valid for MHD turbulence.
			
			One parameter we have assumed has no uncertainty is the kinematic distance of 5.1 kpc \citep{Sanhueza2013}.  The distance uncertainty affects the volume density uncertainty, which is used in the calculation of the B-field strength.  The uncertainty in the distance was estimated as $\sim$10\% \citep{Sanhueza2015}, which contributes $\sim$5\% to the estimated B-field strength. 
			
			A remaining unknown is the inclination angle between the orientation of the cloud long axis and the plane of the sky.  In the calculation of volume density, we assumed zero inclination.  While the true inclination is not known, it is unlikely to be 90$\degr$ (the cloud viewed directly down its spine).  Another unknown is whether the cloud spine curves with respect to the plane of sky, which would result in multiple inclination angles.  For simplicity, the assumption of zero inclination was adopted for the B-field strength estimate.
			
			The correction factor, f, used in Equation \ref{eqnbpos}, also ranges from 0.46 to 0.51 \citep{Ostriker2001}, which introduces an uncertainty of $\sim$10-15\%.  The systematic uncertainty of the PA dispersion is dictated by the uncertainty of the measured PAs, which for Mimir is $\sim$0.6$\degr$ \citep{Clemens2012b}.  This systematic uncertainty is so low that it is usually dominated by the observational uncertainties (a few to $\sim$15\%, as listed in Table \ref{table3}).  
			
			One of the largest uncertainties in the B-field calculation comes from the average volume density.  Since density is an input variable in the CF B-field strength calculation, the uncertainties of B and density are not independent.  Uncertainties will arise in the B-field strength that depend on the method chosen to estimate gas density. Therefore, the method used to estimate the average volume density affects not only the B-field strength, but also any comparison between the B-field, density, and Mass-to-Flux ratio.  The effects of systematic uncertainties of volume density on B-field and \Mphi~uncertainties are discussed in detail below.  In short, we calculated B-field strengths of G28.23 using a variety of average volume density approaches, and found that the B-field strengths ranged by about a factor of 2.  Therefore, we estimate that our systematic uncertainties for the B-field strengths are about a factor of 2 to 3.}
			
			\subsubsection{Systematic Uncertainties due to Volume Density}\label{systematicuncertainties}
			\added{Because the critical density of $^{13}$CO (on the order of 10$^3$ cm$^{-3}$) is larger than the values found for much of the average volume density map, whether $^{13}$CO can become collisionally excited at these densities, and thus be an adequate tracer of the gas motions, becomes an issue.  However, $^{13}$CO is detected across the entire extent of G28.23.  The discrepancy between the critical density of $^{13}$CO and the derived average densities in these regions is likely due to beam dilution, where clumpy, localized regions of higher density are averaged with regions of lower density.  Mapping this substructure would require higher resolution data.  Nevertheless, based on how well the 13$^{CO}$ integrated intensity traces the dust-based column density, the 13$^{CO}$ line widths are likely applicable to calculating the B-field strength in the region.}
			
			\replaced{Some systematic uncertainties exist that must be addressed to more accurately compute B-field properties and compare them to other physical cloud properties.  First, the}{The} average volume density characterizing lines of sight through a cloud can be calculated using different methods \citep[e.g.,][]{Marchwinski2012}.  \replaced{This}{These} will result in different density estimates.  In the present study, the selection of the cloud boundary affects the average volume density estimates.  Our chosen boundary of 50~H$_2$~cm$^{-3}$ might be considered somewhat liberal, whereas selecting a boundary corresponding to the FWHM of the cloud column density may be overly restrictive and would result in ignoring a substantial amount of cloud material and associated B-field.  
			
			To determine the effects of the assumptions made in calculating the average volume densities, we recomputed the B$_{POS}$ strengths and \Mphip~values using average volume densities \replaced{calculated with}{derived from} other boundary values.  Volume density boundaries of 550, 200, and 100~H$_2$~cm$^{-3}$ corresponded to the cloud column density widths equal to the FWHM, \added{$\pm$}2$\sigma$, and \added{$\pm$}3$\sigma$, respectively, where $\sigma$ was derived from fitting the cloud column density \replaced{along}{perpendicular to} the cloud spine, as described in Section \ref{cloud volume density}.  \added{The derived B-field strengths were higher on average for volume densities calculated with stricter cloud boundaries, i.e., for a FWHM boundary.  The B$_{POS}$ strengths derived using the FWHM boundary ranged from $\sim$10--290$\mu$G.}  The best-fitting power law indices of B$_x$ vs. $n$ ranged from 1.24$\pm$0.24 to 0.82$\pm$0.08 for widths equal to the FWHM and \added{$\pm$}3$\sigma$.  The best fits to the relations found by using the FWHM boundary are more uncertain because the number of pixels meeting the criteria to calculate B-field strengths \replaced{decrease}{decreased} drastically (to 15 pixels) compared to the case of the larger boundary out to 50~H$_2$~cm$^{-3}$ \added{(51 pixels)}.  
			
			The use of different boundaries to calculate the average volume densities for G28.23, while changing the indices of the best-fitting power laws to the B vs. $n$ relation, does not change the interpretation that the B-field was not likely to be the dominant force in the formation of the cloud.  The power law indices of the B vs. $n$ relations are still above the 2/3 threshold, which indicate that gravity was likely the dominant force.  This result is still subject to the assumption that the density index for temporal evolution of the B-field matches the density index for present-day structure.
			
			Based on the dependence \replaced{we find for}{of} the B-field strength on density, the B-field morphology, and the \Mphip~values of the cloud, it is \replaced{unlikely that}{unclear whether} the B-field was the dominant force in the formation of G28.23.  \deleted{The relation between \Mphip~and density implies that, at this time, the IRDC is gravitationally dominated in its denser regions.}  
			

	\section{Summary}\label{summary}
			The importance of the B-field in IRDC formation is not \replaced{well-known}{yet understood}.  To determine whether the B-field played a role in the formation of one IRDC, G28.23, we analyzed archival and new NIR polarimetric observations, \added{along with ancillary archival data}.  We posed four questions to investigate the role of B-fields in the formation of G28.23.  One question addressed whether NIR polarizations could probe the B-field in the intermediate layers of clouds as opaque as IRDCs, and three questions addressed the importance of the B-field in cloud formation.  We examined the behavior of NIR polarization percentage with extinction, the relative cloud to B-field PA orientations, the relationship of the B-field strength with density, and finally, the Mass-to-Magnetic Flux ratio across the cloud.
			
			Using a combination of shallow $H$-band and deep $K$-band polarimetric observations, along with dust continuum data from the {\it Herschel} Hi-GAL and ATLASGAL surveys, and spectral line data from GRS $^{13}$CO, we analyzed the properties of the plane-of-sky component of the B-field of G28.23.  We found:
			
			1. The polarization percentages increase as a function of extinction, indicating that our observations do probe the B-field of the outer and intermediate layers \added{(A$_V$ values of 30-40 mag)} of G28.23.
			
			2. The overall relative orientation of the plane-of-sky B-field was neither preferentially perpendicular nor parallel to the projected cloud orientation.  Therefore, it is unclear from the B-field morphology alone whether the B-field affected the formation of the cloud.  The \added{polarization} PAs do, however, show a large-scale pattern.  The distinct relative PA distributions found in different locations around the cloud indicate that the B-field morphology in the region is affected by the presence of the cloud.
			
			3. The dependence of the B-field strength on cloud density was fit with a power law.  The index of the best-fitting power law was 0.73$\pm$0.06, \replaced{which matches}{very similar to} the slope of 2/3 that would imply that the B-field was not dynamically important in the formation of the IRDC.
			
			4. The relation between \Mphip~and density, fit with a power law, indicates that the cloud is magnetically dominated at lower densities and becomes gravitationally dominated at higher densities.  Applying a correction factor to \Mphip~to account for the unknown geometry of the system shows that the \Mphin~vs. density relation crosses unity in the density range $\sim$300--700~H~cm$^{-3}$.  \added{The increase in \Mphip as a function of density implies that the B-field was dynamically important in the cloud's formation, in contrast to the above results.}
			
			Based on the B-field properties found by NIR polarimetry \replaced{, we find that the B-field likely did not strongly influence}{in this study, it is unclear whether the B-field influenced} the formation of IRDC G28.23.  It is likely, though, that the local B-field in the vicinity of G28.23 was influenced by the presence and/or formation of the cloud.

\section{Acknowledgement}
We thank J. Montgomery, T. Hogge, and I. Stephens for constructive discussions on the analysis.  We are grateful to R. Crutcher for permission to include his Zeeman data.  This research was conducted in part using the Mimir instrument, jointly developed at Boston University and Lowell Observatory and supported by NASA, NSF, and the W.M. Keck Foundation.  This research made use of the NASA/IPAC Infrared Science Archive, which is operated by the Jet Propulsion Laboratory, California Institute of Technology (Caltech), under contract with NASA. This publication made use of data products from the Two Micron All Sky Survey, which was a joint project of the University of Massachusetts and the Infrared Processing and Analysis Center/Caltech, funded by NASA and NSF.  This work is based in part on data obtained as part of the UKIRT Infrared Deep Sky Survey.  The ATLASGAL project is a collaboration between the Max-Planck-Gesellschaft, the European Southern Observatory (ESO) and the Universidad de Chile. It includes projects E-181.C-0885, E-078.F-9040(A), M-079.C-9501(A), M-081.C-9501(A) plus Chilean data.  This publication makes use of molecular line data from the Boston University-FCRAO Galactic Ring Survey (GRS). The GRS is a joint project of Boston University and Five College Radio Astronomy Observatory, funded by the National Science Foundation under grants AST-9800334, 0098562, 0100793, 0228993, \& 0507657.  A.E.G acknowledges support from FONDECYT 3150570.  This work was supported under NSF grants AST 09-07790 and 14-12269 and NASA grant NNX15AE51G to Boston University.


\end{document}